\documentclass[fleqn,usenatbib]{mnras}

\usepackage[T1]{fontenc}

\DeclareRobustCommand{\VAN}[3]{#2}
\let\VANthebibliography\thebibliography
\def\thebibliography{\DeclareRobustCommand{\VAN}[3]{##3}\VANthebibliography}

\usepackage{graphicx}	% Including figure files
\usepackage{amsmath}	% Advanced maths commands
\usepackage{amssymb}	% Extra maths symbols
\usepackage{mathtools} 

\usepackage{xcolor}
\usepackage{verbatim}
\usepackage{comment}
\usepackage{changepage}
\usepackage{newtxtext,newtxmath}
\newcommand{\M}{\, \mathcal{M}}
 
\newcommand{\Mpc}{\, \mathrm{Mpc}/h}
\newcommand{\Ml}{\mathcal{M}_l}
\newcommand{\Mg}{\mathcal{M}_g}
\newcommand{\Mgr}{\hat{\mathcal{M}}_g}

\title[The Cosmic Mach Number as an Environment Measure ]{ The Cosmic Mach Number as an Environment Measure for the Underlying Dark Matter Density Field } 

\author[Meriot et al]{
Romain Meriot$^{1}$\thanks{E-mail:  romain.meriot@ens-paris-saclay.fr}
Sadegh Khochfar,$^{2}$, Jose O$\rm{\tilde{n}}$orbe$^{3}$, Britton Smith$^{2}$
\\
$^{1}$Ecole Normale Superieure Paris-Saclay, 4 Avenue des Sciences, 91190 Gif-sur-Yvette, France \\
$^{2}$
Institute for Astronomy, University of Edinburgh, Royal Observatory
, Edinburgh EH9 3HJ, UK \\
$^{3}$
Facultad de Físicas, Universidad de Sevilla, Avda. Reina Mercedes s/n. Campus Reina Mercedes. E-41012, Seville, Spain. \\
}

\date{Accepted 2022 February 7. Received 2022 February 7; in original form 2021 July 29}

\pubyear{2022}

\begin{document}
\label{firstpage}
\pagerange{\pageref{firstpage}--\pageref{lastpage}}
\maketitle

% Abstract of the paper
\begin{abstract}
 Using cosmological dark matter only simulations of a $(1.6$ Gpc$/h)^3$ volume from the Legacy simulation project, we calculate Cosmic Mach Numbers (CMN) and perform a theoretical investigation of their relation with halo properties and features of the density field to gauge their use as an measure of the environment.
 CMNs calculated on individual spheres show  correlations with both the overdensity in a region and  the density gradient in the direction of the bulk flow around that region. To reduce the scatter around the median of these correlations, we introduce a new measure, the rank ordered Cosmic Mach number ($\Mgr$), which shows a tight correlations with the overdensity $\delta=\frac{\rho-\bar{\rho}}{\bar{\rho}}$.  Measures of the large scale density gradient as well as other average properties of the halo population in a region show tight correlations with $\Mgr$ as well. Our results in this first empirical study suggest that $\Mgr$ is an excellent proxy for the underlying density field and  hence environment that can circumvent reliance on number density counts in a region. For scales between $10$ and $100 \Mpc$, Mach numbers calculated using dark matter halos  $(> 10^{12}$ M$_{\odot})$ that would typically host massive galaxies are consistent with theoretical predictions of the linear matter power spectrum at a level of $10\%$ due to non-linear effects of gravity.  At redshifts $z\geq 3$, these deviations disappear. We also quantify errors due to missing large scale modes in simulations. Simulations of box size $\leq 1 $ Gpc/$h$ typically predict CMNs 10-30\% too small on scales of$\sim 100$ Mpc$/h$.  

\end{abstract}

% Select between one and six entries from the list of approved keywords.1
% Don't make up new ones.
\begin{keywords}
Software: simulations -- Large-scale structure of Universe -- Dark matter
\end{keywords}

%%%%%%%%%%%%%%%%%%%%%%%%%%%%%%%%%%%%%%%%%%%%%%%%%%

%%%%%%%%%%%%%%%%% BODY OF PAPER %%%%%%%%%%%%%%%%%%

\section{Introduction}
\label{sec::intro}
\vspace*{0.5cm}
\setcounter{page}{1}
\pagenumbering{arabic}

The current standard paradigm of structure formation assumes a Universe dominated by dark energy, cold dark matter and that is geometrically flat on large scales. The so-called $\Lambda$CDM model has been investigated and confronted with observations over the last several decades and shown to be a remarkably successful \protect\citep{caseLCDM, WMAP9, Planck18}  . 

Within this framework, structures are understood to form in a hierarchical manner, with small structures forming first, and subsequently merging to form larger ones \protect\citep{davishierarchy, bardeenhierarchy}. Simulations have shown the distribution of dark matter and haloes along a complex network of cosmic filaments and nodes  \protect\citep{Springel}.
 The general build up of structure can be analysed  by using dark matter halos/galaxies  as dynamical tracers of the underlying velocity field. A prime example for this is the  pairwise velocity dispersion of galaxies or halos  $\sigma_{12}$. The latter is at the center of the Cosmic Virial Theorem \protect\citep{peebles, CosmicVirialTheorem}, and depends strongly on the local density field. 
Besides the velocity field halos and e.g  their clustering properties via the correlation function \protect\citep{peebles} or cluster counts \protect\citep{clustercosmo} contain information about the background cosmology.

An alternative measure of the the power-spectrum of density fluctuation is the Cosmic Mach Number (CMN) $\mathcal{M}$. It is a dimensionless number first introduced by  \protect\cite{1990ApJ...348..378O}. $\M$ is  defined as the ratio between the bulk flow $u$ and the velocity dispersion $\sigma$ of halos or galaxies, within a given region.  These two quantities, when averaged over a large statistical sample of  regions of the same size allow to estimate the shape of the power spectrum.
 Their ratio, $\M \equiv u/\sigma $, is independent of the normalization of the power spectrum of density fluctuations and of galaxy bias with respect to  the underlying dark matter density field.  Repeated measurements of $\M$ as a function of scale then provide  the shape of the power spectrum. 

\protect\cite{1990ApJ...348..378O} used the CMN to show that the standard Cold Dark Matter (sCDM) model, popular at the time, is inconsistent with observations. Comparisons of Mach numbers between models, simulations, and observations, make it possible to rule out the sCDM with up to $99\%$ of confidence \protect\citep{1992ApJ...395....1S,1993ApJ...408..389S}. Similar results can be expected in the future, as $\M$ can distinguish between other models, such as modified gravity or massive neutrino models \protect\citep{CosmicProbe}.

Although the CMN has first been used as a cosmological  probe, the definition of $\mathcal{M}$ suggests that it depends both on local and large-scale features of the underlying density field. Therefore, it is possible that the Mach number of a single region can act as a measure of the properties of its local environment. In a first attempt  \protect\cite{Moverdensity}  show that the 'local' CMN is a weakly decreasing function of overdensity and galaxy age using hydrodynamical  simulations, demonstrating the use of the  Mach number beyond cosmology and as a potential environment estimator and probe of galaxy formation.  However, the size of the simulation volume in their study $(L=100$ Mpc$/h)^3$ limits the ability to probe high density environments.

The aim of this study is to  revisit and extend the study of the  CMN as an environment tracer   and investigate further correlations with halo properties. Here we will solely focus on the idealised case of dark matter N-body simulations and report empirical findings.   In doing so we introduce a new quantity, the rank ordered CMN, which amplifies existing correlations between the CMN, halo and environmental properties of a region. As we will show in the following, the rank ordered Mach number also reveals a number of tight correlations with properties of the underlying dark matter density field. 

In section \ref{sec::theory}, we give details on the theory behind the CMN and  and lay out the assumptions and approach in  calculating it from our simulations. In section \ref{sec::sims}, we describe the Legacy simulation and the halo catalogs used to calculate $\mathcal{M}$.  Our results are presented in  section \ref{sec:results}. Finally, we discuss our findings and summarise the main conclusions in section \ref{sec::conclusion}.

%-------------------------------------------------------------Cap 2
\section{{Theoretical framework}}
\label{sec::theory}

We here briefly present the Ansatz for the CMN \protect\citep[see for more details e.g.][]{peebles,1990ApJ...348..378O} . The bulk flow $\bold{u(x_0}, r)$ of a sphere of radius $r$ centered at some position $\bold{x_0}$ is given by:

\begin{equation} \label{eq:velocity_real}
    \bold{u(x_0}, r)= \int \mathrm{dx} \, \bold{v(x)} W(\left| \bold{x-x_0} \right|, r)
\end{equation}
Where $W$ is the window function used to average the velocity field inside that region of characteristic size $r$. Natural choices for the window function $W$ are a top-hat or Gaussian, or any function $W(x)$ that is close to $1$ inside of the sphere and quickly decreases towards $0$ as you move away from the region.
Similarly the velocity dispersion of objects in that sphere is given by:
 \begin{equation}
     \sigma^2 (\bold{x_0}, r) = \int \mathrm{dx} \, \left|\bold{v(x)}\right|^2 W(\left| \bold{x-x_0} \right|, r)     -\left|\bold{u(x_0}, r)\right|^2
 \end{equation}

In practice we will be working with discrete data sets composed of individual tracers/halos in which case the following equations will apply:  
\begin{equation}
 \bold{u(x_0}, r)= \sum_{i} \bold{v}_i W_i
\end{equation}
and 
\begin{equation}
  \sigma^2 (\bold{x_0}, r) =\sum_{i} \, \left|\bold{v}_i\right|^2 W_i     -\left|\bold{u}(\bold{x_0}, r)\right|^2
 \end{equation}
In this case, the weights $W_i$ can also represent the uncertainties associated with observational data. Indeed, it seems natural to give less weights to objects with high observational uncertainties. We here assume for $W_i$ a top-hat function.\footnote{ Methods to explicitly calculate weights given selection criteria can be found in e.g. \protect\cite{1993ApJ...408..389S,Agarwal}.}

The Cosmic Mach Number $\M(r)$  is generally defined as  the ratio of the average of $u$ and $\sigma$ \protect\citep{1990ApJ...348..378O}:
\begin{equation}\label{eq:Mg}
\Mg(r) \equiv  \sqrt{\frac{ \left\langle {u^2 (x_0, r)} \right\rangle}{\left\langle {\sigma^2 (x_0, r)} \right\rangle}}
\end{equation}
where the average is taken over all positions $\bold{x_0}$ in space.
This yields a single value of the CMN for a given radius $r$ and describes the average flow of regions of this size. For this reason, we label it the \textit{global} CMN. However, one can also define a \textit{local} CMN, which describes the flow in a single sphere via   :
\begin{equation}\label{eq:Ml}
    \Ml(x_0,r) \equiv  \frac{u (x_0, r)}{\sigma (x_0, r)}
\end{equation}

As we will see below,  $\Mg$ is directly related with the overall background cosmology, and the local $\Ml$  is related to the properties and the environment of an individual given region. We further study and compare these two quantities in section \ref{sec:results}. We use $\mathcal{M}$ to refer to both of these quantities.

In Fourier space, one can rewrite equation \eqref{eq:velocity_real} as: 

\begin{equation}
    \bold{u(x_0}, r)= \int \mathrm{\bold{dk}} \, \bold{\widetilde{v}(k)} \widetilde{W}(kr) e^{-i\bold{k \cdot x_0}}
\end{equation}
Where $\widetilde{W}$   denotes the Fourier transform of the real-space window function $W$, $\bold{\widetilde{v}}$ the transform of the velocity, and $\bold{k}$ the modes.
In the linear regime peculiar velocities and the matter overdensity are related via  \protect\citep[e.g.][]{peebles}: 

\begin{equation}\label{eq:vel_linear}
\bold{\widetilde{v}(k)} = i f H_0 \widetilde{\delta}(k)\frac{\bold{k}}{k^2}
\end{equation}

Where $f$ is the linear velocity growth factor and $H_0$ the Hubble constant.
This in turn implies that the velocity power spectrum $P_v(k)$ is proportional to the matter power spectrum: 
\begin{equation}
    P_v(k) = (f H_0)^2  \frac{P(k)}{k^2}
\end{equation}

 This allows the mean square bulk flow velocity $\left\langle {u^2 (x_0, r)} \right\rangle$ (where the average is taken over all positions $x_0$) to be written as  \protect\citep{Cosmology} : 
 
\begin{align}
\left\langle {u^2 ( r)} \right\rangle = & \int \mathrm{\bold{dk}} \,  P_v(k)  \widetilde{W}^2(kr) \\
= & \int \mathrm{\bold{dk}} \,  (f H_0)^2  \frac{P(k)}{k^2} \widetilde{W}^2(kr)
\end{align}

Since $f\sim \Omega_m^{0.55}$ \protect\citep{Linder2005}, the average bulk flow velocity and velocity dispersion (after a similar derivation) can be written as  :
 
\begin{equation}\label{eq:u2}
    \langle u^2 ( r) \rangle = \frac{H_0 ^2 \Omega_m ^{1.1}}{2 \pi ^2} \int \mathrm{dk} P(k) \widetilde{W} \, ^2 (kr) 
\end{equation}
\begin{equation}\label{eq:sigma2}
    \langle \sigma^2 ( r) \rangle = \frac{H_0 ^2 \Omega_m ^{1.1}}{2 \pi ^2} \int \mathrm{dk} P(k) (1-\widetilde{W} \, ^2 (kr))
\end{equation}
The implications of these equations are relatively straightforward.  $\widetilde{W}$ is close to $1$ on scales larger than $r$, and close to $0$ for modes dominating on scales smaller than $r$. Therefore, the bulk flow velocity mostly depends on the part of the power spectrum that represents large-scale fluctuations, while the velocity dispersion is sensitive to the small-scale fluctuations, given by $k \geq \frac{1}{r}$. Both situations seem to be easily interpreted in real space, the bulk flow of groups of dark matter halos are expected to be dominated by   large scale overdensities, whereas their individual movement diverges from the average one due to local fluctuations in the density field. \\
It is interesting to note that equations \eqref{eq:u2} and \eqref{eq:sigma2} imply that the global Mach number does not depend on cosmological parameters such as $H_0$ and $\Omega_m$, nor on the normalization of the power spectrum. Thanks to this, $\Mg$ as a function of $r$ is essentially a measure of the shape of the power spectrum on scales where the  linear theory still applies.

In the non-linear regime, the relation between the power spectrum and the Mach number is not as straightforward, as equation \eqref{eq:vel_linear} does not hold strictly.  To focus only on linear fluctuations the density field is generally smoothed on  scales of $r_s$ large enough (typically chosen to be $\sim 5$ Mpc) to remove non-linear fluctuations \protect\citep{1990ApJ...348..378O}.  A common choice is to use a k-space window function that is a Gaussian  (\protect\cite{1990ApJ...348..378O}):

\begin{equation}
  \widetilde{W}= e^{-\frac{(kr_s)^2}{2}}  
\end{equation}

We use the above equation when calculating CMNs from both theoretical power spectra and power spectra computed from the simulations. When working with catalogs of discrete objects (i.e., galaxies or halos), we divide each region into cubes of length, $r_s$, and calculate the value of any property as the mass-weighted average within the cube. For mass, we simply take the sum of virial masses. After this, we calculate the values of $\sigma$ and $u$ as the unweighted average of all cubes. In Appendix A, we examine the effect of using unweighted averages instead of mass-weighted averages for properties within each cube.

\section{The Legacy simulations}

\label{sec::sims}
\begin{table*}%[h!]
    \centering
    \begin{tabular}{|c|c|c|c|c|c|c|c|c|}
         \hline
         $\Omega_L$ &$\Omega_m$ &$\Omega_b$ &h &$\sigma_8$ &$n_S$ &$Y_{He}$ &$N_{\mu}$ &$w$  \\
         \hline
         $0.7150$ &$0.2850$ &$0.0450$ &$0.6950$ &$0.828$ &$0.9632$ &$0.2480$ &$3.0400$ &$-1.0 $ \\
         \hline
    \end{tabular}
   \protect\caption{WMAP9 Cosmological parameters used in the Legacy simulations.}
    \label{cosmoparameters}
\end{table*}

  The Legacy project  consist of a suite of cosmological N-body, dark matter (DM) only simulations run with the Gadget-4 code\footnote{The repository of the code can be found here: https://gitlab.mpcdf.mpg.de/vrs/gadget4} (\protect\cite{gadget}, \protect\cite{gadget4}). Initial conditions were generated using the MUSIC code (\protect\cite{music}). The cosmological parameters used in the simulations are listed in Table \ref{cosmoparameters}. They were chosen to be consistent with the WMAP9 results (\protect\cite{WMAP9}).  \\

 In this work we will be using the main Legacy simulation which consist of a simulation volume of side length $1600 \Mpc$ and $2048^3$ DM particles of $5.43 \times 10^{10} \, \mathrm{M}_{\odot}$ each. Gravity is softened on a scale of 5.3328 \rm{kpc}/h. The simulation has been run from $z=99.0$ to $z=0$, and $101$ snapshots with the full particle information are stored at regular intervals,  evenly spaced in log of the expansion factor. On top of this halo catalogs are generated for each snapshot using the Rockstar halo finder \protect\citep{rockstar}\footnote{The repository of the version can be accessed here: https://bitbucket.org/gfcstanford/rockstar/src/main/ }. The volume and resolution of the simulation ensures to includes both small and large modes that are essential for the calculation of the CMN, as we will discuss later.
 
  \begin{figure}
    \centering
    \begin{adjustwidth}{-0.9cm}{-0.0cm}
     \includegraphics[scale=0.59]{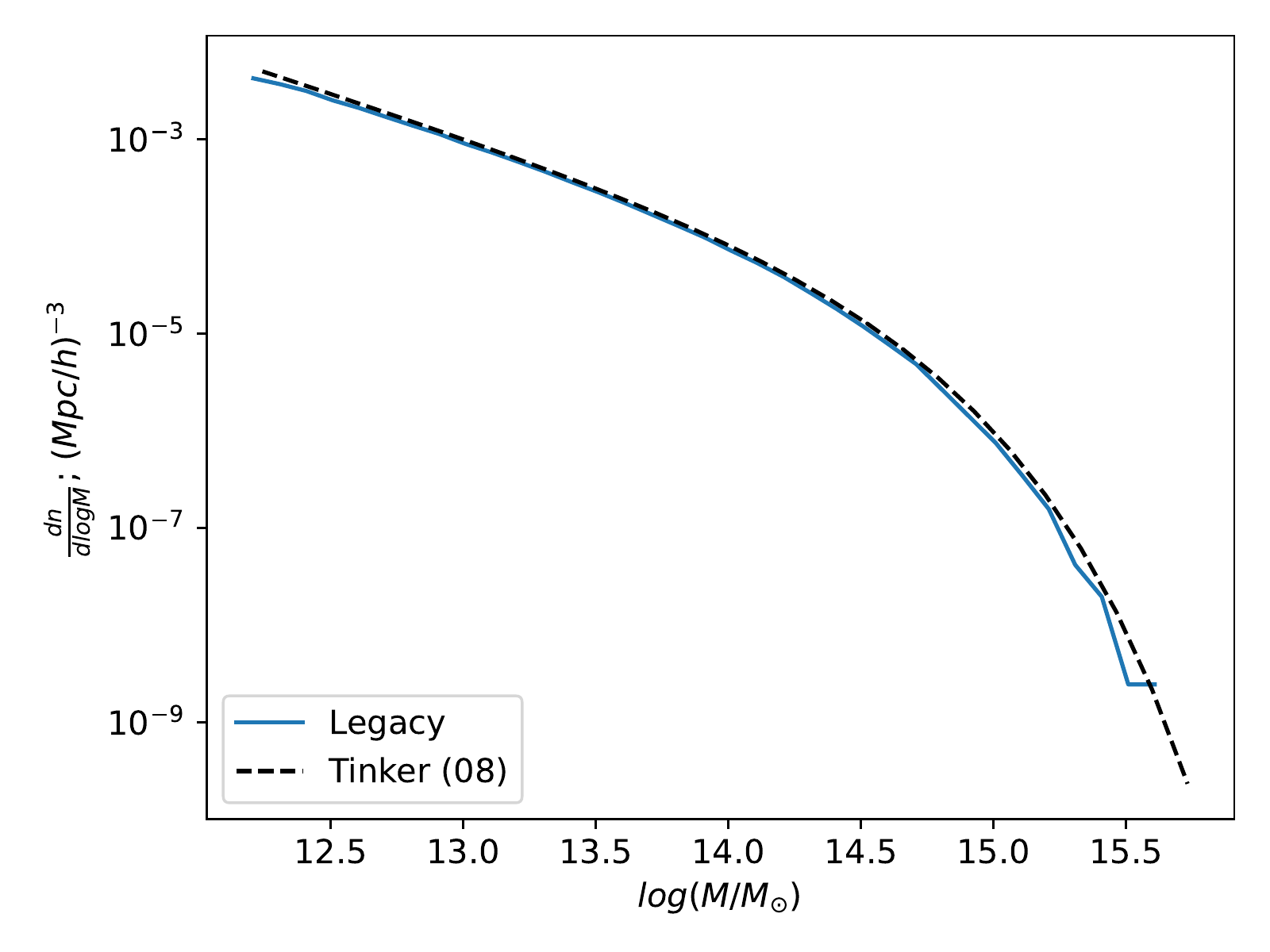}
     \end{adjustwidth}
    \protect\caption{Mass function of dark matter halos in the Legacy simulation at redshift z=0. The mass function is consistent with the mass function fit presented in \protect\cite{TinkerMF}. All halos contain at least 20 dark matter particles.  }
    \label{fig:sims}
\end{figure}

 In Figure \ref{fig:sims} we shows the mass functions of the Legacy simulations used in this study. Over a mass range from $10^{12.2}-10^{15.5}$ M$_{\odot}$ the mass function is in good agreement with the fit proposed by \protect\cite{TinkerMF}. The latter has been matched against various different simulations by other groups and suggests that our simulations are consistent. In the following we will only consider halos with particle numbers $n \ge 20$ ( $\ge 1.086 \times 10^{12} $ M$_{\odot}$) which corresponds to the mass range shown in the figure.

%\chapter{Methods and results}
\section{Results}
\label{sec:results}

We start this section by presenting results that confirm that the computation of the CMN from our simulations is consistent with the expected trends, before extending the analysis to estimate environmental dependencies.
\subsection{Power Spectrum}

\begin{figure}%[h!]
\begin{adjustwidth}{-1cm}{-1.5cm}
    \centering
    \includegraphics[scale=0.58]{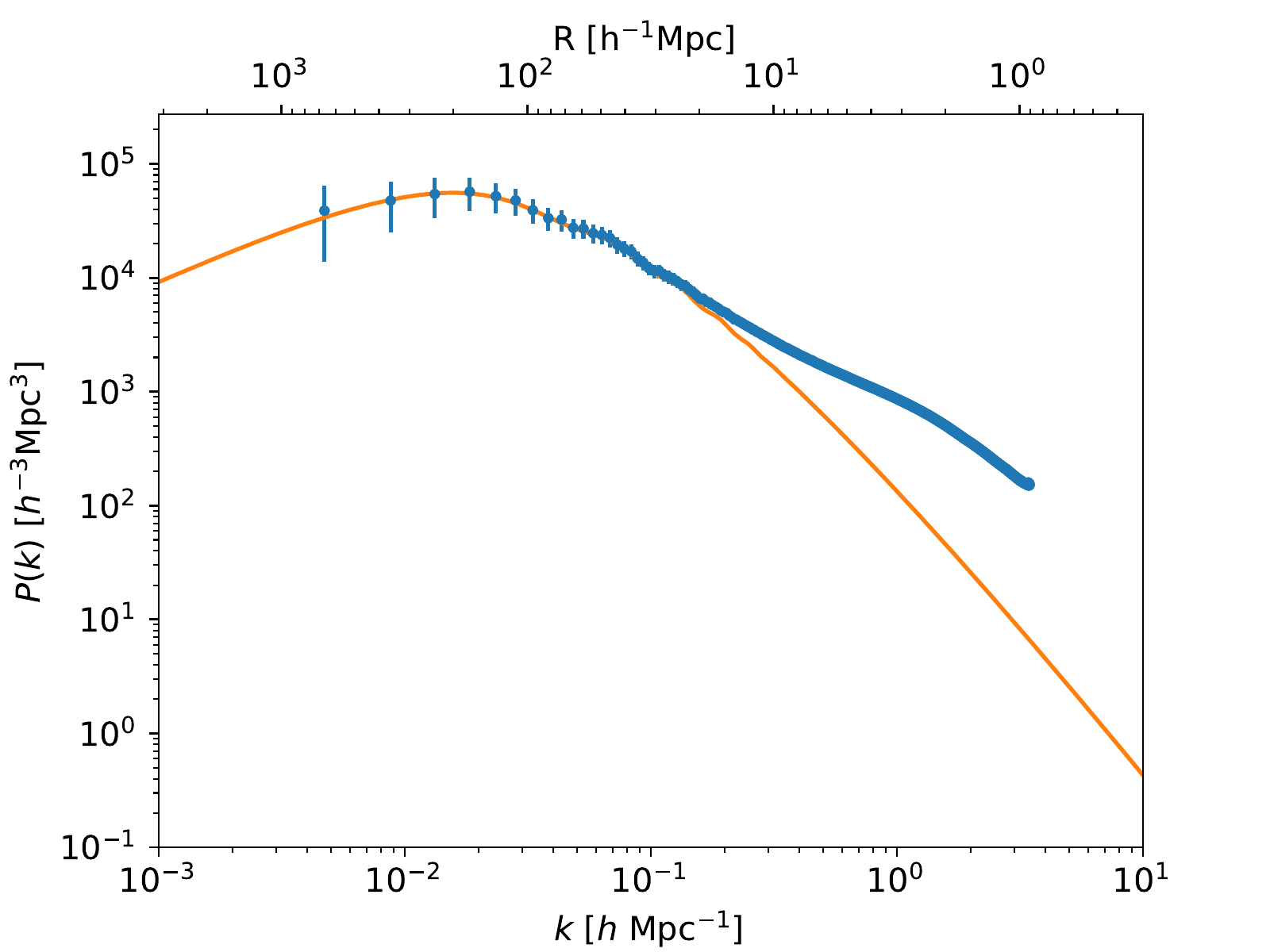}
    \end{adjustwidth}
   \protect\caption{Input power spectrum used to generate the initial conditions for the simulation   linearly extrapolated to $z=0$ (orange), and the power spectrum calculated from the simulation at $z=0$ (blue). Error bars show $1-\sigma$ errors. Deviations on small scales ($\le 5 \Mpc$, large $k$) are due to non-linear growth of structure. Overall the measured spectrum is in good agreement with the expectations from the input spectrum within errors.    }
    \label{powerspectrum}
\end{figure}

The CMN is directly related to the shape of the power spectrum via Eqs. \eqref{eq:sigma2} and \eqref{eq:u2}.        
In Figure \ref{powerspectrum} we show the input power spectrum from which the initial conditions of the simulations are generated, linearly extrapolated to $z=0$. The power spectrum measured from the simulation data at $z=0$ with the nbodykit software \protect\citep{nbodykit} is shown in the same figure as symbols with error bars.  the errors are calculated using the estimator  presented in \protect\cite{errorbarspk1} and \protect\cite{errorbarspk2} via: 
\begin{equation}
    \left(\frac{\Delta P}{P(k)}\right)^2 = \frac{1}{C(k)}\left[1 + \frac{2}{N_p P(k)} + \frac{1}{N_p^2 P^2(k)} \right]
\end{equation}
Where $C(k)$ is the number of statistically independent available wavenumbers, and $N_p$ the numbers of particles.
 The power spectrum of the simulation is in very good agreement with linear predictions for scales $\ge 5$ Mpc$/h$, but drops slightly, although still consistent within the errors,  below the expectations at scales close to the simulation box sizes $> 1000 $ Mpc$/h$. On scales $< 5 $ Mpc$/h$ clear deviations from the linear power spectrum are found which are due to non-linear growth of structures. The derivation of the CMN starting from Eqs. \eqref{eq:vel_linear} assumes linear growth of structure and hence is most applicable to the linear and mildly non-linear range of the power spectrum. We will therefore focus on CMNs on larger scales and smooth over scales  of 5 Mpc$/h$ in the following sections.      
 
\subsection{The Global Cosmic Mach Number $\Mg$}

\subsubsection{Simulation Requirements} 
{
\begin{figure*}
{\begin{adjustwidth}{-2.4cm}{-2.cm}
    \centering
    \includegraphics[scale=0.66]{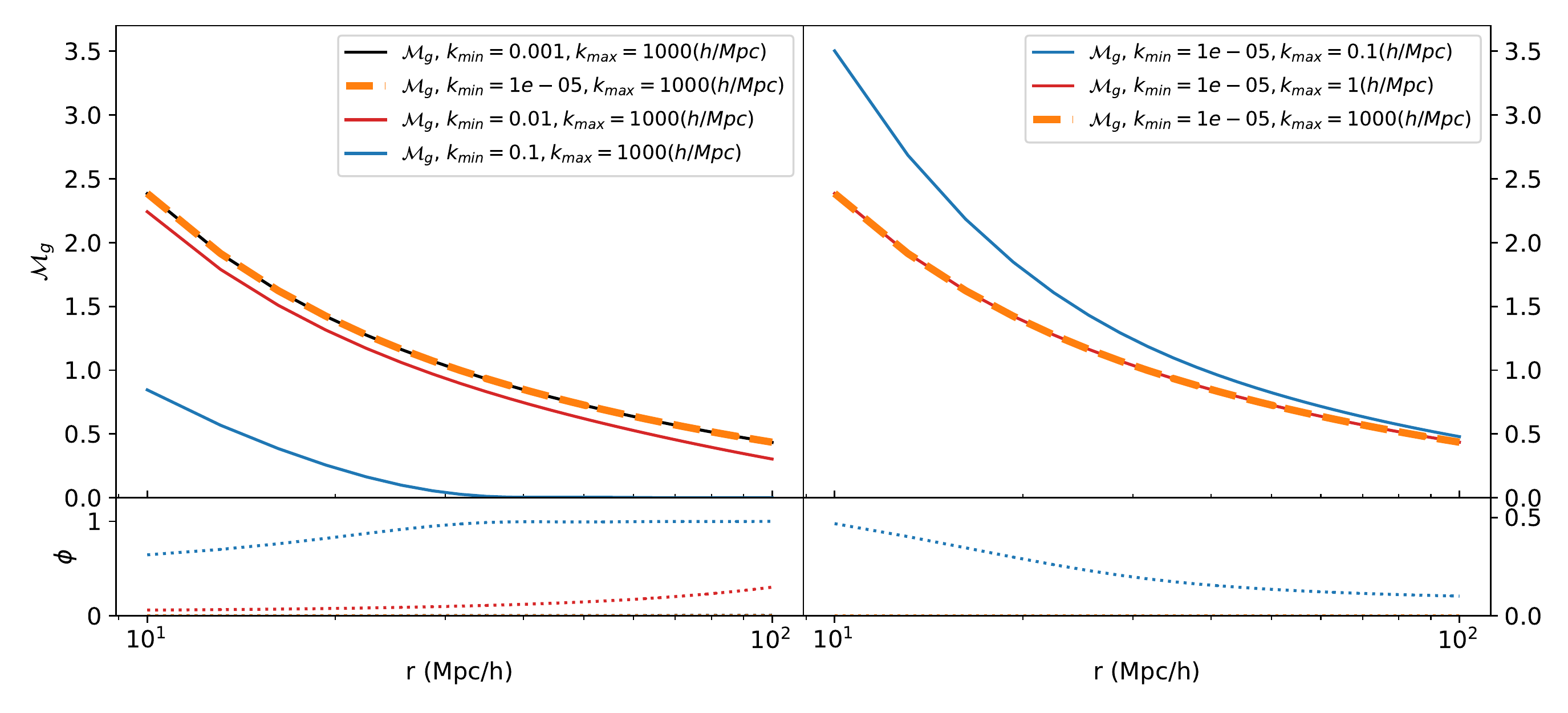}
    \end{adjustwidth}
    }
{ { \protect\caption{ \protect\small $\Mg$ as a function of the radius, for different upper and lower limits of the integration range of the power spectrum. This effectively truncates the power spectrum on one side or the other representing simulation volume and resolution limitations.  \textit{Left:} $\Mg$ as a function of $r$ for different $k_{min}$. $k_{max}$ has been set at $1000\, \mathrm{h/Mpc}$ . \textit{Right:} $\Mg$ as a function of $r$ for different $k_{max}$. $k_{min}$ has been set, $10^{-5} \, h/\mathrm{Mpc}$. The solid/dashed lines are the results for different ranges of modes. 
The dotted lines in the bottom panel show the relative difference, compared to the case taking the full power spectrum  from $k_{max}$ to $k_{min}$ corresponding to all modes on scales between $\sim 5 \Mpc$ and $\sim 10^3 \Mpc$.  The black line, corresponding to $k_{min} = 10^{-3}\sim h/$Mpc, is identical to the orange one, which confirms that, in $\Lambda$CDM, 
for $k_{min} \le 10^{-3} \sim h/$Mpc results for the CMNs converge. The bottom panel shows the residual
${ \phi= \frac{ \mathcal{M}_{\mathrm{full}} - \mathcal{M}_{\mathrm{truncated}}}{\mathcal{M}_\mathrm{full} } }$
, the relative error between the results from the initial full spectrum (orange dashed line) and those from the truncated spectra.
} }}
    \label{Mvsk}
\end{figure*}
}
In this section we will quantify the intrinsic errors in the calculation of the global Mach number depending on the simulation setup and show that the Legacy simulation used here minimises such errors.
 The CMN depends on the power spectrum across many scales, but all scales do not contain the same power. As numerical simulations cover the power spectrum only on scales from the Nyquist frequency of the simulation setup up to the size of the simulation volume, one expects this to affect CMNs measured in simulations. Too coarse resolution will result in an underestimate of the velocity dispersion $\sigma$ due to missing small scale power, and a simulation volume not covering to large enough scales will result in an underestimate of the bulk flow $u$. Figure \ref{Mvsk} shows the global Mach numbers as predicted analytically using different integration limits $k_{\mathrm{min}}$ and $k_{\mathrm{max}}$ for the theoretical input power spectrum, mimicking different simulation setups. As expected, increasing the lower limit $k_{\mathrm{min}}$, and therefore ignoring large-scale modes, result in lower Mach numbers, as these large modes contribute to the bulk flow. Eventually, the Mach number will tend towards zero, once $k_{\mathrm{min}} \sim 1/r$  corresponding to the size of the region over which the CMN is calculated.  Similarly, lowering $k_{\mathrm{max}}$ results in higher Mach numbers, as fewer small scales are resolved and contribute to the velocity dispersion. 
 However, the relation between the Mach number and the power spectrum strictly only holds in the linear regime, so increasing the resolution will not change the Mach numbers measured at redshift zero, as the contributions on scales smaller than $\sim 5 \Mpc$ are smoothed over in our approach. A higher resolution might be useful when measuring $\mathcal{M}$ at a larger redshift with a smaller smoothing scales, when  non-linearities are still confined to small scales.  
 Based on Fig. \ref{Mvsk} constraints on CMNs with accuracy better than $10\%$ are only achieved for simulations that include modes up to scales of $1000 \Mpc$, and at least resolve all scales in the linear regime  (i.e. larger than $\sim 5 \Mpc$). Mach numbers computed with the full input power spectrum from $k_{\mathrm{min}}= 10^{-5}\, h/\mathrm{Mpc}$ to $k_{\mathrm{max}}= 1000\, h/\mathrm{Mpc}$ are identical to those calculated with parameters consistent with the  full box size of the Legacy simulation, supporting the notion that our simulation volume in this study is large enough to capture all relevant scales and to predict  accurate Mach numbers. Present state-of-the-art cosmological hydrodynamics simulations like e.g.  The Eagle \protect\citep{eagle1, eagle2} or Illustris \protect\citep{TNG} use volumes of a few hundred $\Mpc$ on the side. Therefore, one would  expect CMNs calculated using these simulations to be $\sim 5\%$ smaller than what they should be on scales of $\sim 10 \Mpc$, and $\sim 30 \%$ smaller at scales of a $\sim 100 \Mpc$.

\subsubsection{The CMN as a function of scale}

We here show the distribution of global CMN, calculated as the average CMN of 50 randomly placed non-overlapping  spheres in the simulation volume using Eq. \eqref{eq:Mg}. \footnote{Our tests indicate that result converge for 50 or more spheres. }

\begin{figure}
\begin{adjustwidth}{-0.5cm}{-0.0cm}
    \centering
      \includegraphics[scale=0.58]{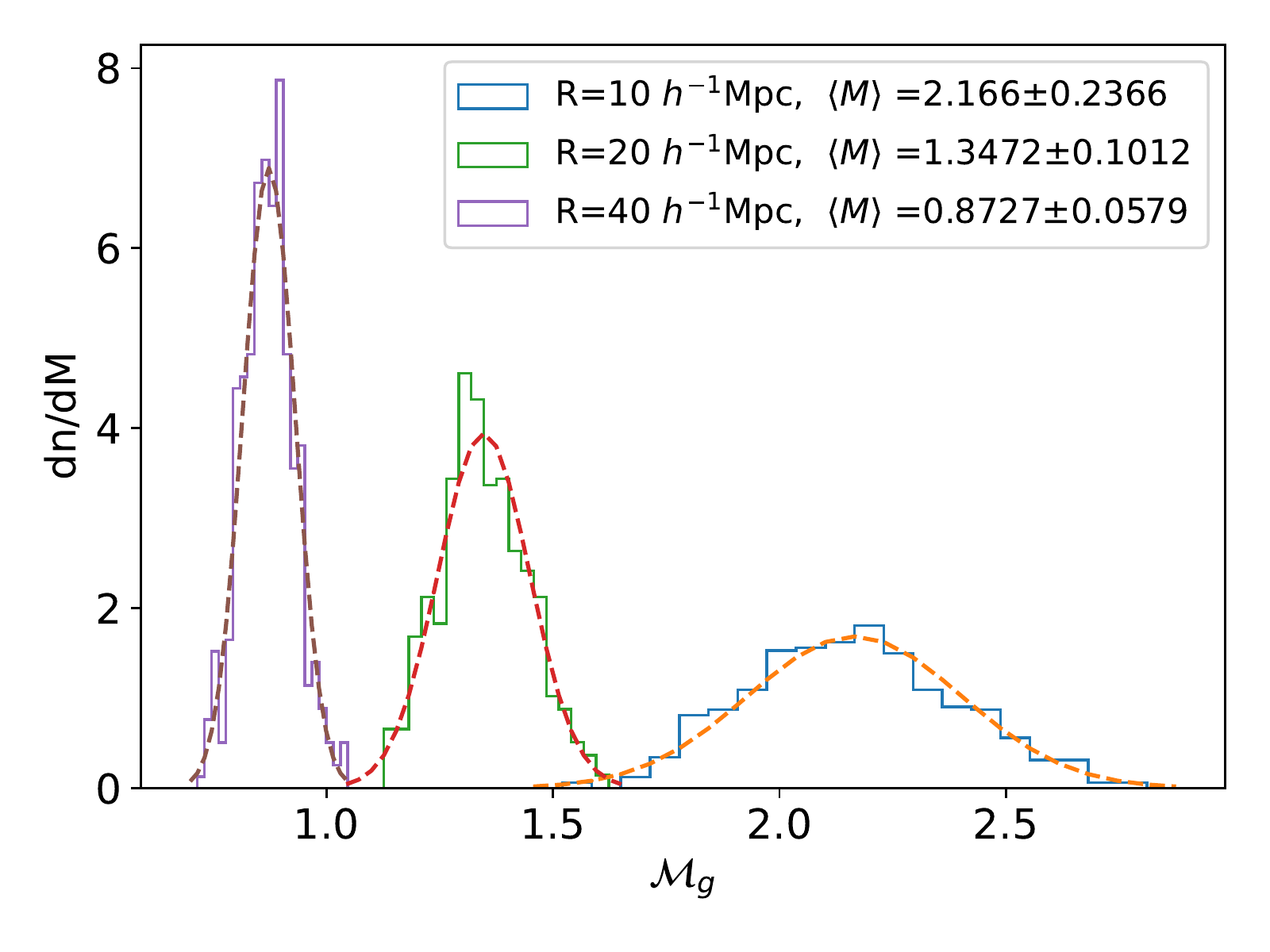}
     \end{adjustwidth}
   \protect\caption{\small{Distribution of the CMN for spheres of radius $10$, $20$, and $40 \Mpc$. The distributions are well fitted by Gaussian distributions (dashed lines).}}
    \label{Mdist10}
\end{figure}

Figure \ref{Mdist10} shows the distribution of 500 global CMNs, calculated each averaging over 50 spheres of radius $R=10, \, 20 \,$ and $ 40  \Mpc$, respectively. The distributions are well described by  Gaussians. As the radius increases, both the average and width of the distributions decrease reflecting the shape of the power spectrum and the increase in homogeneity of the universe averaging over larger scales, respectively.

\begin{figure}
\begin{adjustwidth}{-0.5cm}{0cm}
    \centering
    \includegraphics[scale=0.72]{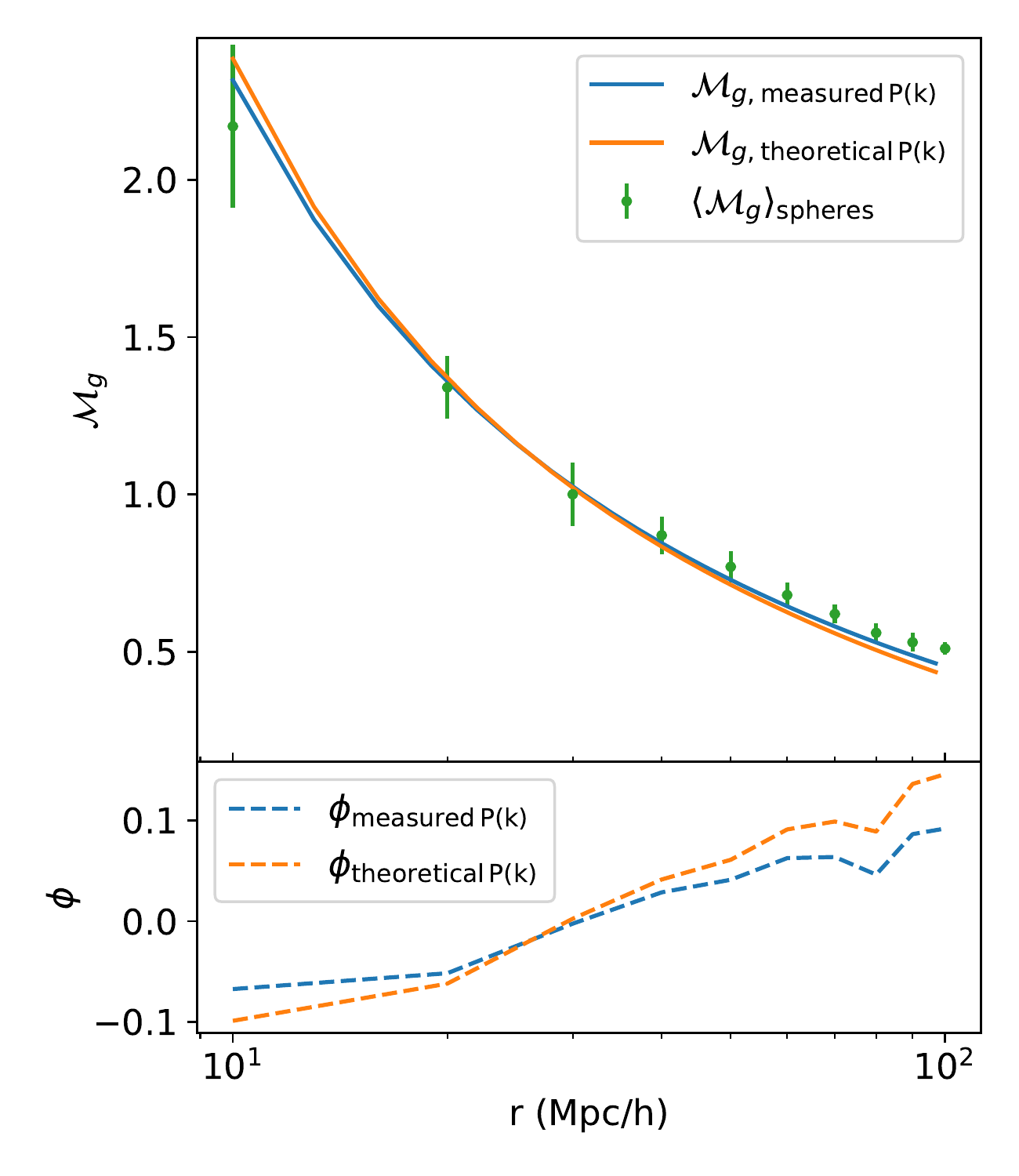}
    \end{adjustwidth}
   \protect\caption{\small{The global CMN as a function of radius. The orange and blue solid lines show the results inferred from the theoretical input and measured power spectra, respectively,  integrating Eqs. \eqref{eq:sigma2} and \eqref{eq:u2}. The green points are calculated using the halos in the simulation (100 groups of 50 spheres at each radius). All results were obtained using a $5 \Mpc$ smoothing length.} Error bars are standard deviations around the mean value. The bottom panel shows  the relative error between the results from the spectra and those from the spheres, e.g $\phi_{\mathrm{measured \, P(k)}}= \frac{ \langle \mathcal{M}_{g} \rangle _{\mathrm{spheres}} - \mathcal{M}_{g,\mathrm{measured \, P(k)}}}{\mathcal{M}_{g,\mathrm{spheres}} } $ 
       } 
    \label{MvsR}
\end{figure}

 Figure  \ref{MvsR} shows the predicted CMN from the simulation for radii $R=10-100 \Mpc$.
In order to compare with the theoretical expectations, CMNs were calculated using Equations \eqref{eq:u2} and \eqref{eq:sigma2} by integrating the extrapolated input power spectra shown in Figure \ref{powerspectrum}, with smoothing applied over $r_s = 5 \Mpc$ for both simulation data and theoretical predictions, as discussed in section \ref{sec::theory}. The power spectrum was smoothed using a Gaussian filter and the simulation data using cubes, but as noted by \protect\cite{1992ApJ...395....1S} and also seen in our own tests, the actual shape of the smoothing function is not as important as the smoothing scale. 

The orange solid line of Figure \ref{MvsR} shows CMNs obtained by integrating the theoretical power spectrum, while the blue line shows the results integrating the measured power spectrum (the orange and blue spectra of Figure \ref{powerspectrum}, respectively ) from $r=10$ to $100 \Mpc$ in Eq. \eqref{eq:u2} and \eqref{eq:sigma2}. The green points are  obtained by calculating the average $ \left<\Mg \right>$ over  a hundred $\Mg$. The latter are averaged over $50$ spheres at each radius. Error bars are the standard deviation around the average $ \left<\Mg \right>$. There is a small disagreement between the results using the theoretical input spectrum versus the measured one, which is a consequence in the  slight difference in the power spectra shown in  Figure \ref{powerspectrum},  which lies within the errors of the measured spectrum.  At small scales, $\Mg$ obtained from averaging spheres match the predictions quite well, but as scales increases, a small discrepancy starts to appear, as the measured average value $ \left<\Mg \right>$ (green points) lie systematically above the expected values from integrating the power spectra. 

We do not expect this deviation to be due to the smoothing scale. We confirm this in Appendix \ref{app:smooth} by changing the smoothing scales $r_s$ systematically and finding the same deviation.
A possible explanation for the deviation is that linear theory underestimates the bulk flow of structure as seen in the  velocities of clusters \protect\citep{underestimatevel2,underestimatevel1,underestimatevel3}. 
Further support for the notion of this being related to the build-up of large scale structure in the cosmic web comes in section \ref{z_mach} in which we show that these deviations  on large scales are systematically reduced going to higher redshift when structures on these scales are still not well developed. 
Deviations are also expected to some extent given that halos are not perfect tracers of the underlying matter density field. As we show in Appendix \ref{app:smooth} our results improve if all dark matter particles within a region are taken into account. However, we will continue to calculate CMNs using halos in  this study and not dark matter particles to allow the connection to observations which rely on galaxies whose movements follow closely those of their hosting dark matter halos.  
In general, the agreement between all three methods shown in Fig. \ref{MvsR}  is good, with a maximum deviation of the measured $\left<\Mg \right>$ at scales of $\sim 100 \Mpc$ that leads to an error of $\sim 10\%$. As presented in section \ref{z_mach}, our results indicate that this deviation disappears at higher redshift, when non-linearities on small large scale are not developed yet.

\begin{figure}
\begin{adjustwidth}{-2.2cm}{-2cm}
        \centering
        \includegraphics[scale=0.72]{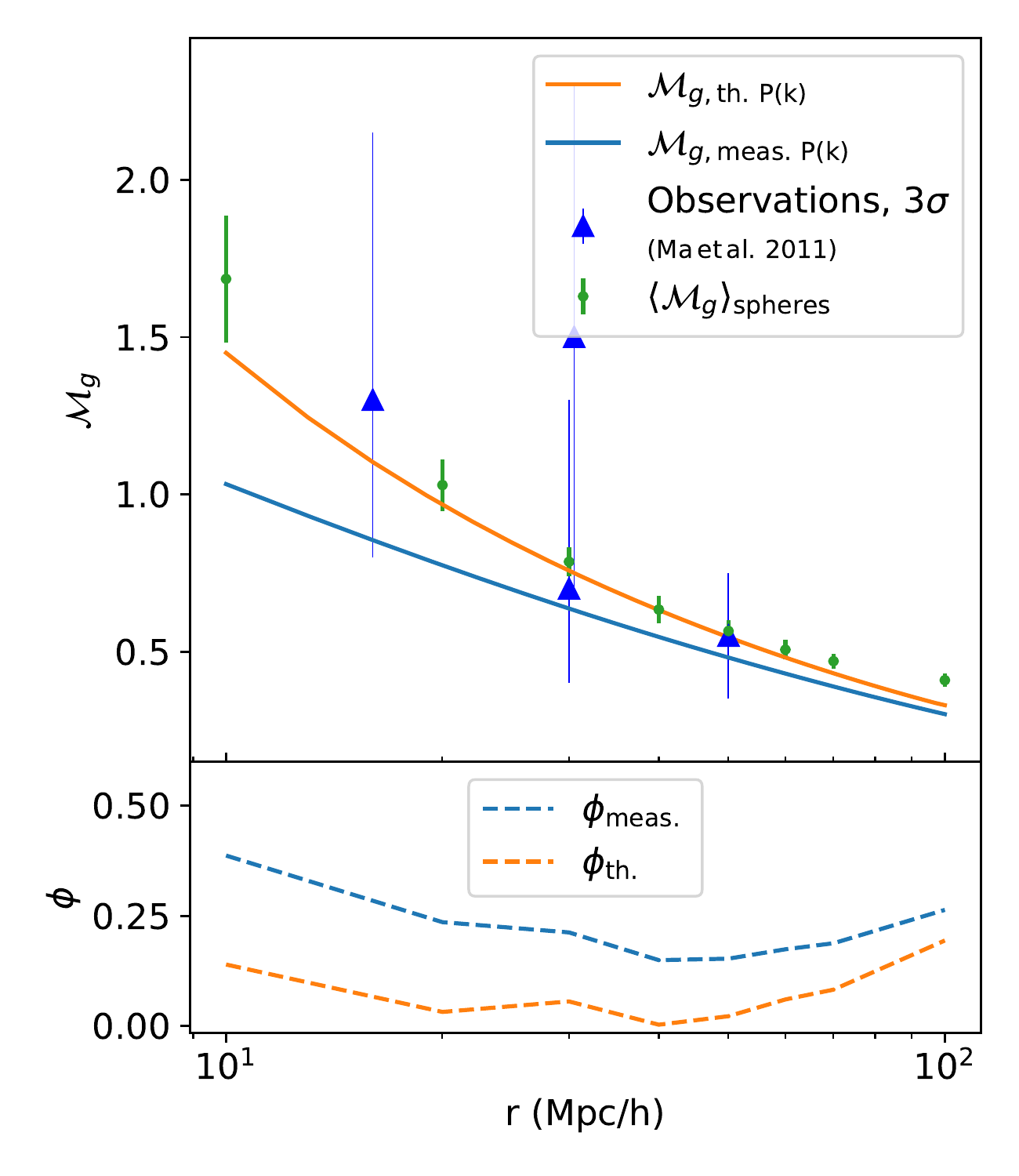}
        \end{adjustwidth}
        \protect\caption{  $\Mg$  vs $r$ without smoothing on scales smaller than $r_s= 5$ Mpc$/h$. Coincidentally, the simulation results overlap with calculations using the linear input power spectrum of the simulation (orange line), while they are offset with regards to the calculations using the actual power spectrum of the simulation (blue line).  The blue triangles represent the results from observations (obtained without smoothing) presented in \protect\cite{CosmicProbe}. Like on the previous figure, the residuals on the bottom panel are calculated via
        e.g $\phi_{\mathrm{meas. \, P(k)}}= \frac{  \langle \mathcal{M}_{g} \rangle _{\mathrm{spheres}} - \mathcal{M}_{g,\mathrm{meas. \, P(k)}}}{\mathcal{M}_{g,\mathrm{spheres}} } $  We show that the CMNs calculated using halos are consistent with observations, but they are significantly larger than the results obtained from the measured spectrum, highlighting the need to smooth over non-linearities.  
         }
\label{nosmoothing}
\end{figure}

\begin{figure}
    \begin{adjustwidth}{-0.6cm}{0cm}
    \centering
    \includegraphics[scale=0.7]{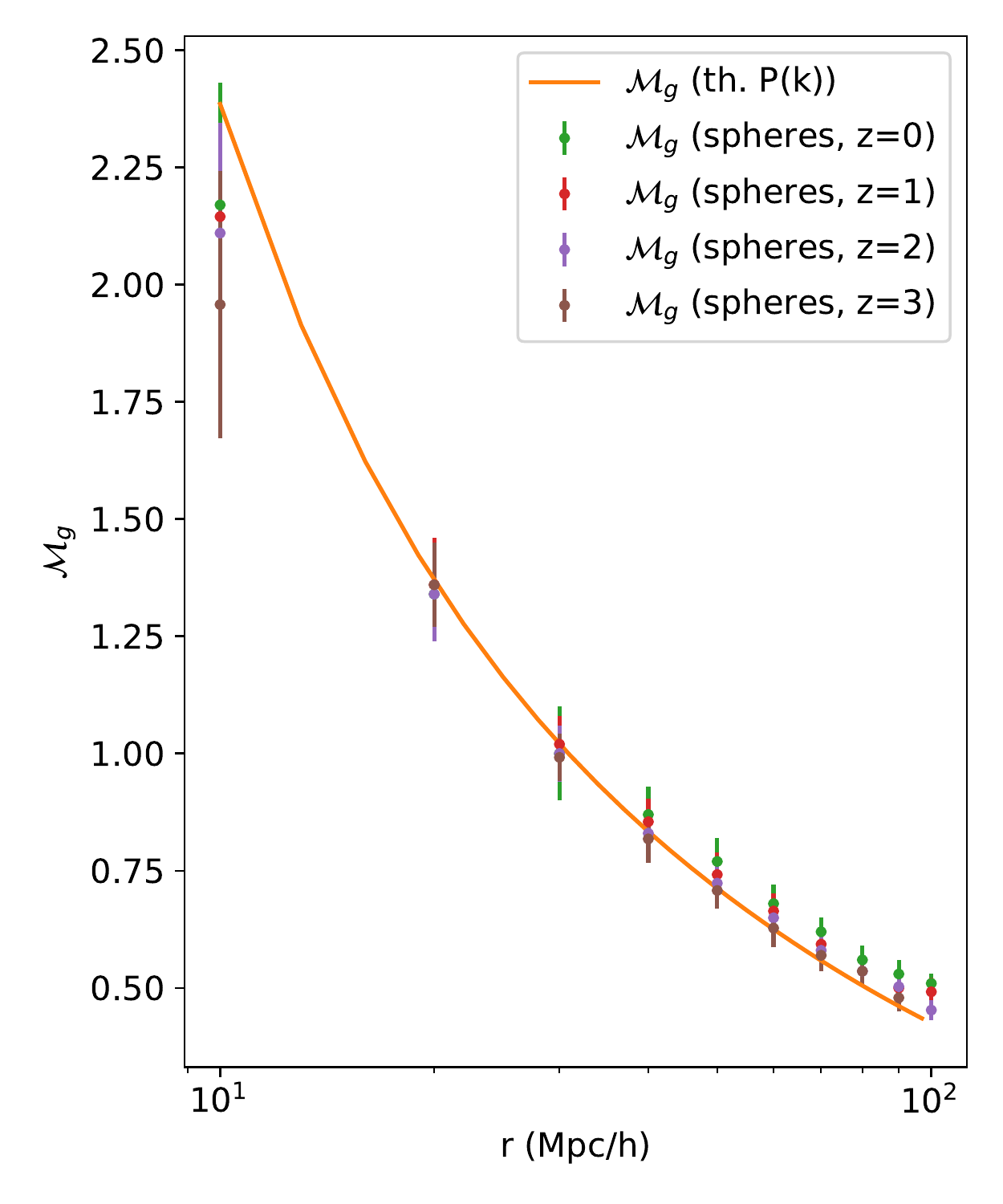}
    \end{adjustwidth}
   \protect\caption{$\Mg$ as a function of radius, for redshift 0 to 3. High redshift points are consistent with the theoretical spectrum, as non linearities develop later. All results were obtained using a $5 \Mpc$ smoothing length.}
    \label{Mvsz}
\end{figure}

\subsubsection{Impact of non-linearities } 

 \begin{figure*}
 \begin{adjustwidth}{-0.5cm}{0cm}
    \centering
    \includegraphics[scale=0.52]{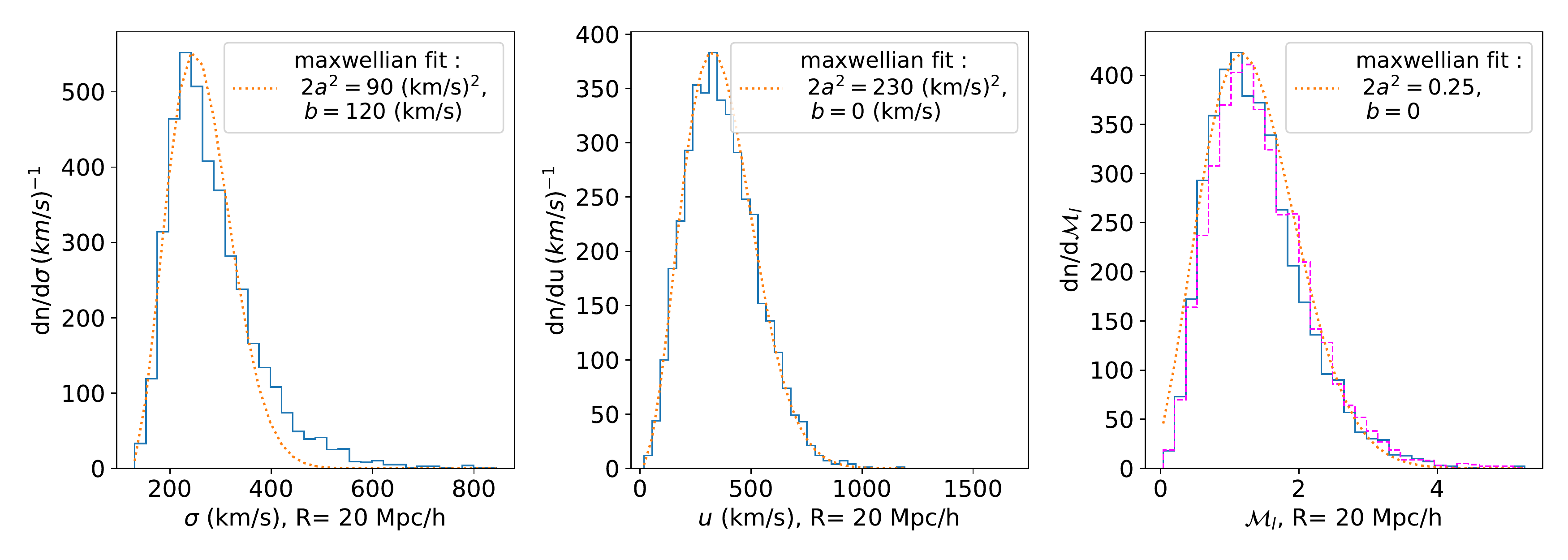}
         \end{adjustwidth}

   \protect\caption{Distributions of $u$,$\sigma$, and $\mathcal{M}$, for 4000 realizations, fitted with distributions of the form $ P(x) = \sqrt{\frac{2}{\pi}} \frac{(x-b)^2 e^{(x-b)^2/2a^2 }}{a^3}$.The magenta dashed histogram in the third panel shows the results of a Monte-Carlo simulation of the CMN based on randomly choosing values for  $u$ and $\sigma$ from the fitted distribution shown in the two left most panels.  } 
    \label{Mudisp}
    \end{figure*}

The main assumption in the derivation of the CMN is that linear theory holds on scales of interest. While it should do so also in the mildly non-linear regime, it is interesting to gauge the impact of non-linear structure on the estimates of the CMN on different scales.    As we have shown in Fig. \ref{powerspectrum} the power spectrum deviates on scales $<5 ~$Mpc$/h$ from the linear extrapolation. As a consequence smoothing is applied over these scales  in simulations as well as in  observations   \protect\citep{1990ApJ...348..378O}.  
Figure \ref{nosmoothing}  shows our results if we do not apply any smoothing in the calculation of $\Mg $ for the same  three methods we applied in the previous section,  i.e. the CMNs calculated with the theoretical and measured spectra, and using the halos.  The Mach numbers calculated with the measured spectrum (including non-linearities) using linear theory lie below  those found with the theoretical input spectrum linearly extrapolated to $z=0$, as the non-linearities cause an increase in power on small scales and result in a generally larger velocity dispersion $\sigma$. The Mach numbers calculated using our simulated halos deviate from the other estimates.The maximum deviation appears on scales of $\sim 10 \Mpc$ where there is a $\sim 18 \%$  difference between the results of the halos and the results of the theoretical input spectrum.  This increases to $\sim 40 \%$ for the measured spectrum.  On large scales, the results from the halos and the theoretical input spectrum overlap and are off-set compared to those of the measured spectrum, which includes significant excess power due to non-linearities. 
The agreement between $\Mg$ derived from halos with the theoretical one is most likely coincidence, as linear theory, which connects the CMN calculated in real space with the spectrum integration, only holds if the non-linearities are not too large .
These results confirm  that extending linear calculations for $\Mg$ to the non-linear regime is not as straightforward as simply using the full non-linear power spectrum. For completeness, we also compare $\Mg$ obtained from halos in the simulation to observations (both without smoothing) presented in \protect\cite{CosmicProbe}, which  reveals good agreement within albeit large uncertainties of the observations. Future surveys will be key to significantly reduce uncertainties in observations and provide stronger constraining power.

 \begin{figure*}
    \begin{adjustwidth}{-5cm}{-5cm}
    \centering
    \includegraphics[scale=0.62]{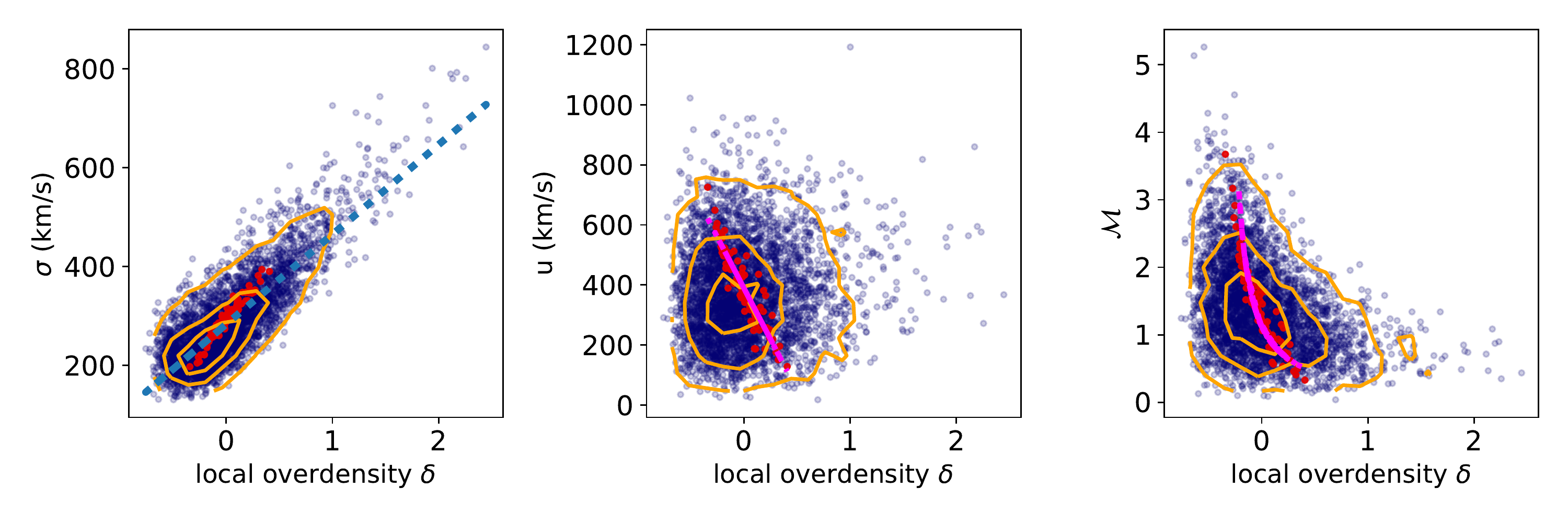}
        \end{adjustwidth}
   \protect\caption{\small{\textit{Blue points}: local $\sigma$, $u$, and $\Ml$ for individual halos, as functions of the overdensity $\delta$ enclosed in the corresponding spheres, for 4000 random spheres of radius 20 Mpc$/h$. The contours show the regions where the distribution lie above $5\%$, \, $33\%$, and $66\%$ of its maximum, respectively. $\sigma$ is correlated with $\delta$, while $u$ does not show any strong trend, as a consequence  $\mathcal{M}$ is a decreasing function of the overdensity. The blue dashed line shows the best linear fit for $\sigma$. \textit{Red points}: Same for the rank ordered global Mach numbers $\Mgr$, averaged over groups of 50 spheres of similar local Mach numbers $\Ml$. The averaged bulk flow $u$ now depends clearly on the averaged overdensity $\delta$ while no such trend is visible in the local values (blue points).The magenta dotted lines shows the best fit for $\Mgr$ with the form $\Mgr = \frac{1}{\alpha + \beta \delta}$, with $\alpha=0.895$ and $\beta=2.68$ , as well as the best linear fit for the global ranked bulk flow $u = A \delta + B$ with $A = -680, B = 389$.  } }
    \label{Mudispvsrho}
\end{figure*}

\begin{figure*}
\begin{adjustwidth}{-5cm}{-5cm}
    \centering
    \includegraphics[scale=0.85]{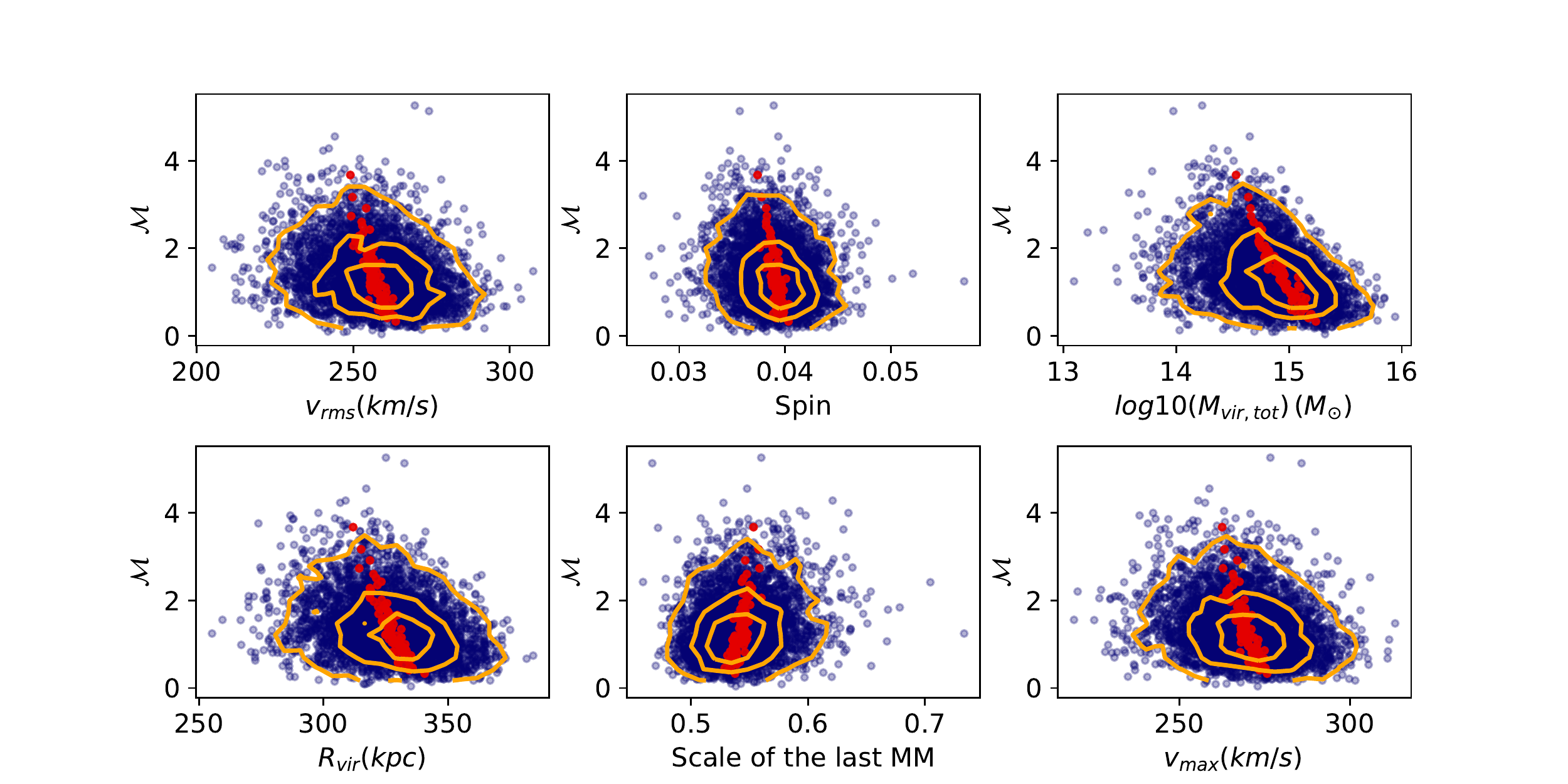}
    \end{adjustwidth}
   \protect\caption{  $\Ml$ (blue points) as a function of average halo properties of halos contained in spheres of radius $20 ~ {\mathrm Mpc}/h$. The studied properties are: spin, virial radius $R_{vir}$, velocity dispersion $v_{rms}$, maximum circular velocity $v_{max}$, scale of the last major merger. $M_l$ shows a broadening of the the total Mass distribution going to smaller values. The rank ordered global Mach number$\Mgr$ (red points) shows tight correlations with halo properties  (see section \ref{rank} for more details on $\Mgr$).  }
    \label{Mvsprop}
\end{figure*}

\subsubsection{Evolution with redshift}\label{z_mach}

The global CMN by construction does not show any redshift dependency, as both $u$ and $\sigma$ scale with the power spectrum similarly see Eq. \eqref{eq:u2} and \eqref{eq:sigma2}.
If anything, the Mach numbers calculated with dark matter halos should match more accurately the theoretical predictions at higher redshift, when  non-linearities have not appeared yet.
The scale at which non-linearities kick in decreases with redshift  thus suggesting that the CMN provides a  more accurate measure of the power spectrum at smaller scales at high redshift.

 In  Figure \ref{Mvsz} we show $\Mg$ as a function of radius for different redshift $z$ as well as the results obtained doing the theoretical calculation from the corresponding power spectra, measured in the simulation. First, we see that the results at $z>0$ agree with those at $z=0$. This is the case both for the results measured  from  spheres in the simulation volume, and for those obtained by integrating the power spectrum. We also notice that the agreement between the measured $\Mg$ and the theoretical predictions does get better  at higher redshift.  The differences between the results at $z=0$ and those at $z>0$ come solely from the non-linearities that appear as redshift decreases.

\subsection{The local Cosmic Mach Number $\Ml$ }

As shown on Figure \ref{Mdist10}, the distributions of the CMNs are fairly wide. Given the relation of $\Mg$ with the matter power spectrum, which describes the density fluctuations of the Universe, the difference between the Mach numbers at a given scale are expected to arise from differences in the "local" environment on scales over which halos are averaged. In the following sections, we investigate potential links between the Mach number of a single region and the properties of the region over which we average including their environment. The Mach numbers $\Mg$ presented so far are obtained from averaging over a large number of randomly chosen non-overlapping spheres on a given scale $r$, in effect averaging  over variations in, $u$ and $\sigma$, and their dependence on physical properties. To investigate physical connections more directly we calculate an alternative Mach number which does not rely on ensemble averaging. We will calculate the \textit{local Mach number} $\Ml$   for individual spheres as the ratio of the bulk flow and the velocity dispersion of halos inside said spheres in contrast to the global Mach number $\Mg$  which is computed using many such spheres. Note that $\Mg \equiv \left< u \right> /\left< \sigma \right> $  is by construction not the average of $\Ml= u/\sigma$ for individual spheres. 
The \textit{local} nature of  $\Ml$ trades the clear connection to the power spectrum of the simulation for, potentially, more straightforward relations to environmental physical properties. In the following,  $\Ml$ will be explored.

Figure \ref{Mudisp}  shows the distributions of the bulk flow $u$, the velocity dispersion $\sigma$, and the local Mach number $\Ml$, for one thousand spheres of radius  $20 \Mpc$ \footnote{ We  use a radius of $20 \Mpc$ as our fiducial value because it is small enough to guarantee efficient computing and large enough to be not affected by the  smoothing over $ 5 \Mpc$. We note that our results show the same trends for e.g. larger spheres of $50 \Mpc$ radius. }.
The bulk flow is well described by a Maxwellian distribution. Each component of the average velocity of the group of halos in a sphere follows a Gaussian distribution  so the bulk flow of the group follows a Maxwellian distribution \protect\citep{Kumar2015GravitationalPW}. 
The velocity dispersion does not follow a Maxwellian distribution as closely as the bulk flow, as it shows some excess at the large $\sigma$ tail, but the discrepancy is not large. The velocity dispersion is however closer to a Maxwellian distribution than the results shown by \protect\cite{Moverdensity}. The difference can be attributed to the smoothing of non-linear small-scale fluctuations which has been omitted in \protect\cite{Moverdensity}. 
 The Mach number is well described by a Maxwellian distribution too, in agreement with previous findings \protect\citep{Moverdensity}. Randomly choosing  $u$ and $\sigma$ values using their respective distributions and calculating $\Ml$ produce the green dashed histogram in the third panel, which agrees very well with the measured distribution of $\Ml$ in the simulation. This confirms that $u$ and $\sigma$ can be considered independent variables which is a feature of Gaussian random density fields, where modes on different scales are independent of each other.

Since the definition of  $\Ml$ relies only on dynamical properties of a system of halos, and dark matter dynamics are primarily driven by the the density field, we expect the Mach number to show correlations with density estimates of the local environment and its vicinity.
 Figure \ref{Mudispvsrho} shows for 4000 spheres of radius $20  \Mpc$ $\Ml$, $u$, and $\sigma$, against the overdensity $\delta = \frac{\rho - \rho_m}{\rho_m}$, with $\rho$ as the density in a sphere and $\rho_m$ the average density in the Universe.  The density $\rho$ is calculated using all the dark matter particles in the relevant region, not only those associated with dark matter halos.\footnote{The overdensities presented on this figure reach values as high as $\sim2$, as scales of a few dozens of $\Mpc$ show mild non-linear features. However, we note that the calculated Mach numbers are still consistent with linear theory on these scales (as shown on figure \ref{MvsR}) and that the empirical results presented here remain unchanged in strictly linear contexts, be it at larger radii or higher redshifts.}
While the bulk flow shows no clear correlation with the overdensity, the velocity dispersion seems to be linearly related with it. As a result, $\Ml$ is a decreasing function of overdensity. 

\begin{figure*}
\begin{adjustwidth}{-5cm}{-5cm}
    \centering
    \includegraphics[scale=0.62]{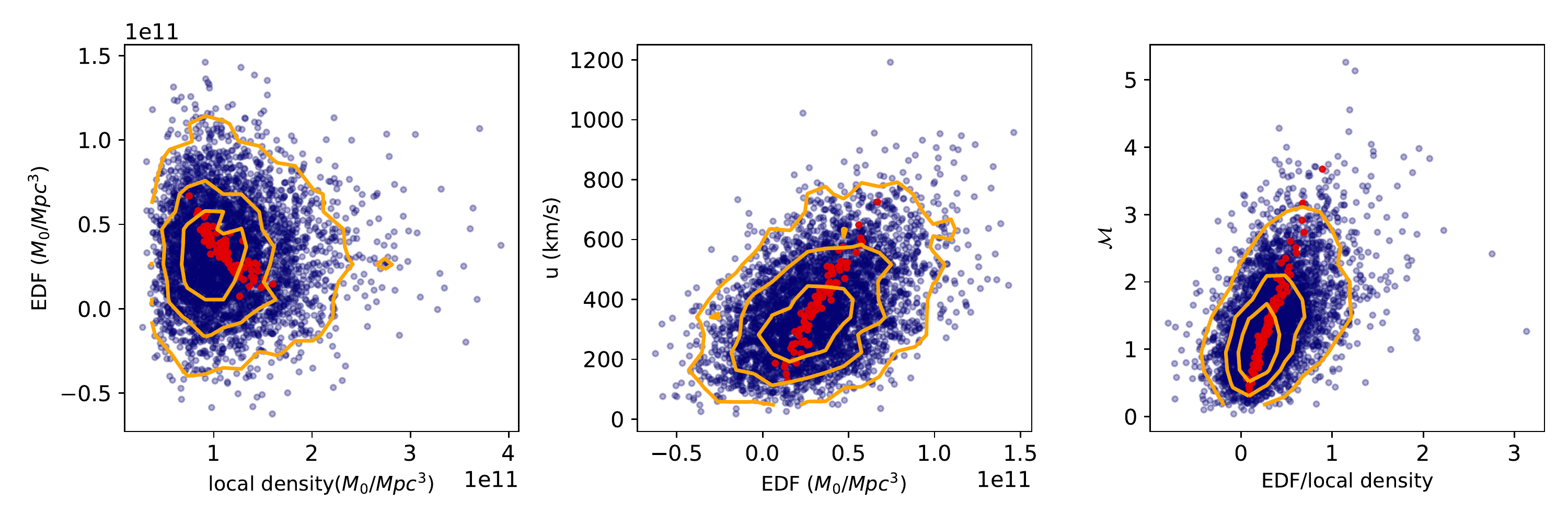}
    \end{adjustwidth}
   \protect\caption{\textit{Left}: Environmental density fluctuation (EDF) between two nearby, diametrically opposed $60 \Mpc$ spheres vs local density in a 20 $\Mpc$ sphere, for 4000 realizations. \textit{Center:} $u$ vs EDF, the panel shows a positive correlation, which is expected due the relation between $u$ and the density gradient in the region. \textit{Right}: $\Ml$ vs the ratio of EDF to local density. $u$ depends on the EDF, which is a measure of the large scale  density gradient. The EDF and $\delta$ are uncorrelated, so $\Ml$ is also a measure of their ratio. Blue shows local values, while red is for rank ordered global CMN $\Mgr$ obtained from averaging in groups of spheres of similar local Mach numbers (see section \ref{rank} for more details). In addition to the relations with local values, we observe a strong correlation between $\delta$ and the EDF, which causes the relation between the global $u$ and $\delta$ seen in Figure \ref{Mudispvsrho}. }
    \label{allvsrhos}
\end{figure*}

 The pair-wise velocity dispersion $\sigma_{12} (r)$, which is the dispersion of a pair of tracers separated by a distance $r$ shows a very similar trend. The Cosmic Virial Theorem  predicts  a relation between $\sigma_{12}$ and density, and this relation has been previously observed in N-body simulations \protect\citep{pairwise_sigma}.  It is therefore not surprising that the velocity dispersion of the groups exhibit a similar behavior. These results are overall consistent with those presented in \protect\cite{Moverdensity}, even though the Legacy simulation includes significantly more large scale modes that are essential for the calculation of $u$, $\mathcal{M}$, and the overdensity, as shown in Figure \ref{Mvsk}. We also note that we do not find a clear correlation between $\delta$ and the bulk flow. As for $\Ml$ we find that the median of the distribution of overdensities $\delta$ systematically gets larger as  $\Ml$ gets smaller with an increasing scatter. Our results predict that regions with $\Ml \le 1 $ have on average a matter densities corresponding to $\delta > 0 $. The latter is not surprising given the positive correlation of $\sigma$ with $\delta$ and the lack of a clear correlation of the bulk flow, but it also reflects the fact that more overdense regions are typically the dominant large scale gravitational sources in their environment and therefore do not show much bulk flow themselves but have mostly matter moving  towards them. Halos in underdense regions like voids in contrast could either be moving with a large bulk velocity towards a more overdense region and hence show large $\Ml$ or still be in the process of getting accelerated. 

\begin{figure}%[h!]
\begin{adjustwidth}{-1cm}{-0cm}
    \centering
    \includegraphics[scale=0.6]{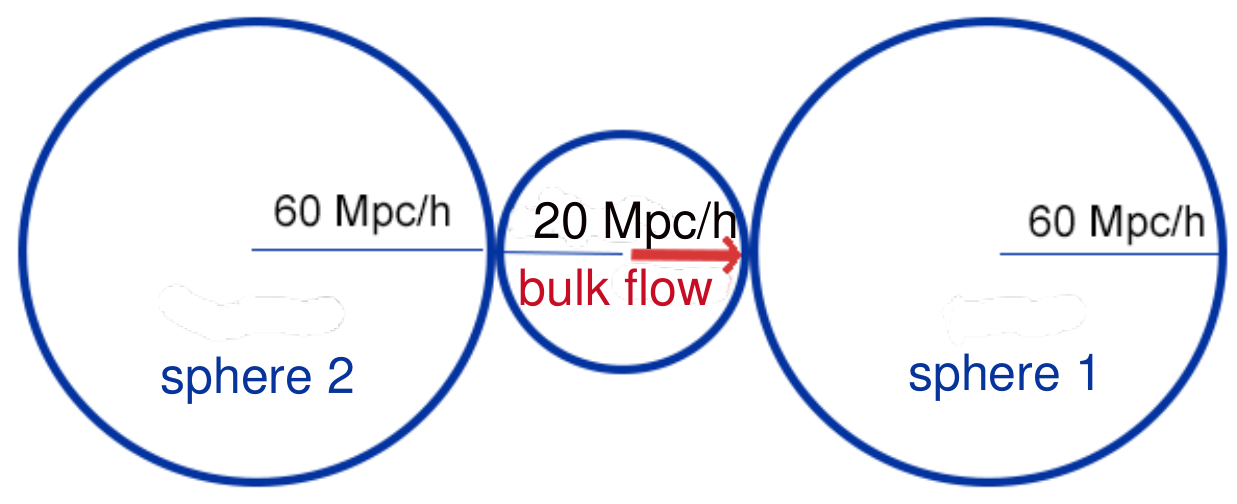}
\end{adjustwidth}
   \protect\caption{Definition of the environmental density fluctuation. Sphere 1 (2) is in the direction (opposite direction) of the bulk flow. The EDF is defined as the density difference between these two large spheres via $ \eta_{{\mathrm EDF}}= \rho_1 - \rho_2 $.}
    \label{EDF}
\end{figure}

Figure \ref{Mvsprop} shows $\Ml$ plotted against the average halo properties of the objects in 4000 spheres of radius $20 \Mpc$, except for the virial mass $M_{\mathrm{vir}}$, which is the total of all halos,
for the following properties: Spin, virial radius $R_{\mathrm{vir}}$, velocity dispersion $v_{\mathrm{rms}}$, maximum circular velocity $v_{\mathrm{max}}$, scale of the last major merger and spin 
\footnote{Note that the results presented in this paper do not change significantly if the mass weighting is replaced by a number-weighted average as show in Appendix \ref{app:smooth}}.

$\Ml$ does not show any strong correlation with average halo properties, except for a weak correlation with the total mass of halos, which is a tracer of the underlying  density field. We find similar trends also for the average  velocity dispersion $v_{rms}$ and maximum circular velocity $v_{max}$, both of which are related to the total mass of halos in a region. Interestingly, the average last major merger halos in high $\Ml$ regions experience is slightly later than in low $\Ml$ regions although these regions can have individually last major mergers that happen later. The intrinsic scatter of the time of the average last major merger decreases with $\Ml$. The average spin parameter of halos show a very slight dependence on $\Ml$ in becoming somewhat smaller for larger $\Ml$.

\begin{figure*}
\centering
\begin{adjustwidth}{3.cm}{0cm}
    \includegraphics[scale=0.8]{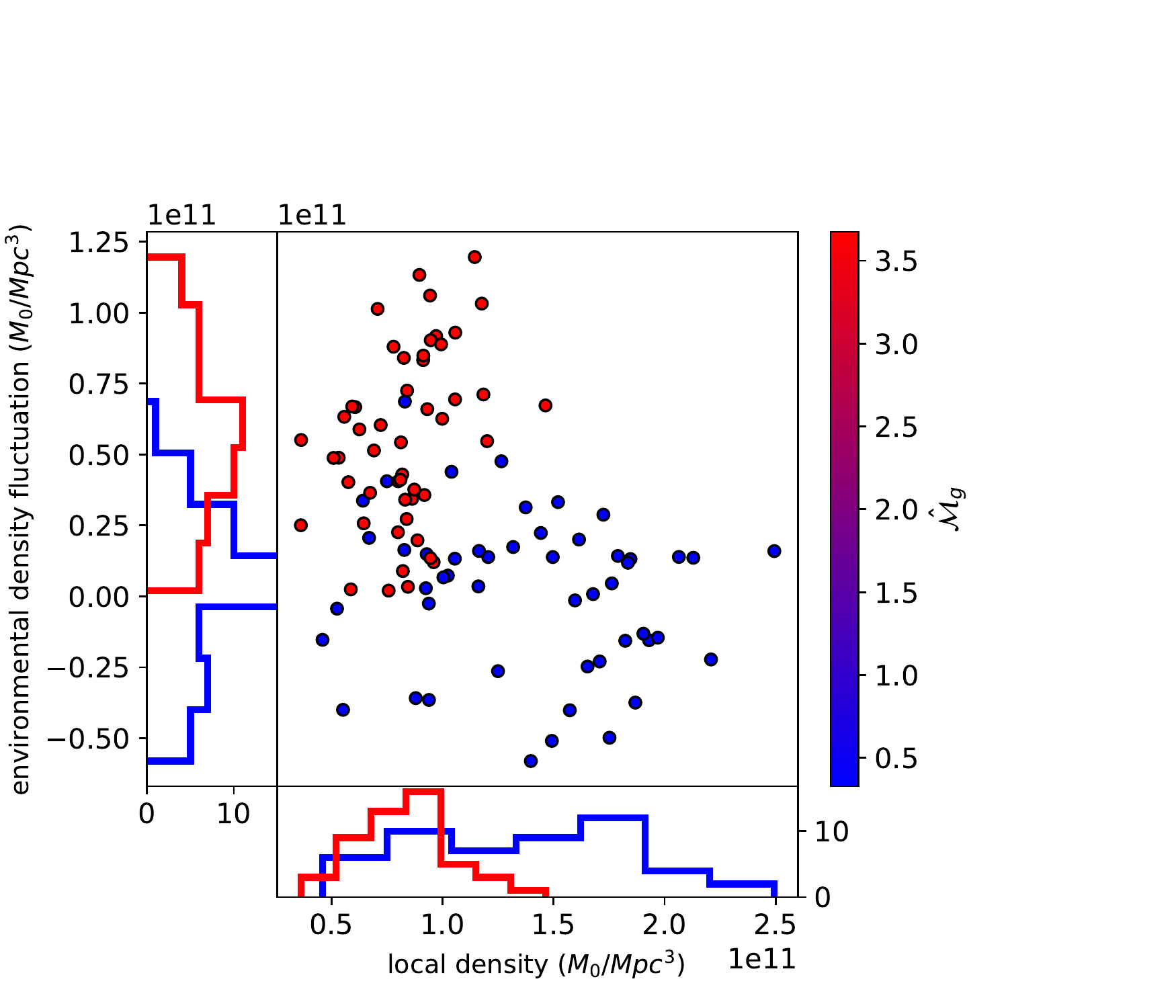}
\end{adjustwidth}
   \protect\caption{  \protect\footnotesize 
    \textit{Main panel}: same as the top left panel of \protect\ref{allvsrhos} but showing the extreme groups (\textit{red :} highest $\Mgr$; \textit{blue :} lowest)  \textit{Side panels}: histograms of the density and EDF of each group. }
    \protect\label{rhoscolor}
\end{figure*}

The expectation is that a large scale bulk-flow of a region follows density fluctuations around that region that source gradients in the gravitational potential.   
To quantify this effect  we calculate the density in two larger adjacent spheres, one in the direction of the bulk flow $\rho_1$, and one in the opposite direction $\rho_2$ (see Figure \ref{EDF}). We call the density difference between these two spheres $\eta_{\mathrm{EDF}}=\rho_1-\rho_2$ the  \textit{environmental density fluctuation} (EDF). The top right  panel of Figure \ref{allvsrhos} shows how the bulk flow depends on the environmental density fluctuation.
A positive correlation is clearly visible. This is explained by the fact that a large overdensity will attract our group of halos if its influence is not compensated by an overdensity in the opposite direction. However, the scatter is quite large\footnote{ This does not depend on the choice of sizes for the adjacent spheres. We have tested this for spheres of $40 \Mpc$ and $80 \Mpc$ finding similar results. } 
($\sigma_{{\mathrm EDF}}/ \left\langle \eta_{{\mathrm EDF}} \right\rangle   \sim 3 $ for the $33\%$
contour),  which can be explained by two factors. First, a single overdensity in a nearby region is only the simplest scenario, the two spheres are but a rough simplification of the often complex topology of the gravitational potential which drives the motion of a group of halos. Also, the environmental density fluctuation is only an approximation of the local density gradient in the direction of the bulk flow, and this gradient has been shown to strongly correlate with the bulk flow \protect\citep{Kumar2015GravitationalPW}.  This  large  scatter results in cases of negative EDF for $\Ml$.

Since the velocity dispersion linearly depends on the local density, the local Mach number correlates tightly with the ratio of the environmental density fluctuation and the local density, as shown on the  panel all to the right of Figure \ref{allvsrhos}. The pair of large spheres do not overlap with the central small sphere in order to avoid any accidental correlation between the environmental density and the local density, as shown in Figure \ref{EDF}. Overall, Figure \ref{allvsrhos} shows a clear relation between the Mach number and the properties of the density field of that region.Alternative definitions of the local density gradient by \protect\cite{Kumar2015GravitationalPW} show similar results.

 \begin{figure}
\begin{adjustwidth}{-1cm}{0cm}
    \centering
     \includegraphics[scale=0.59]{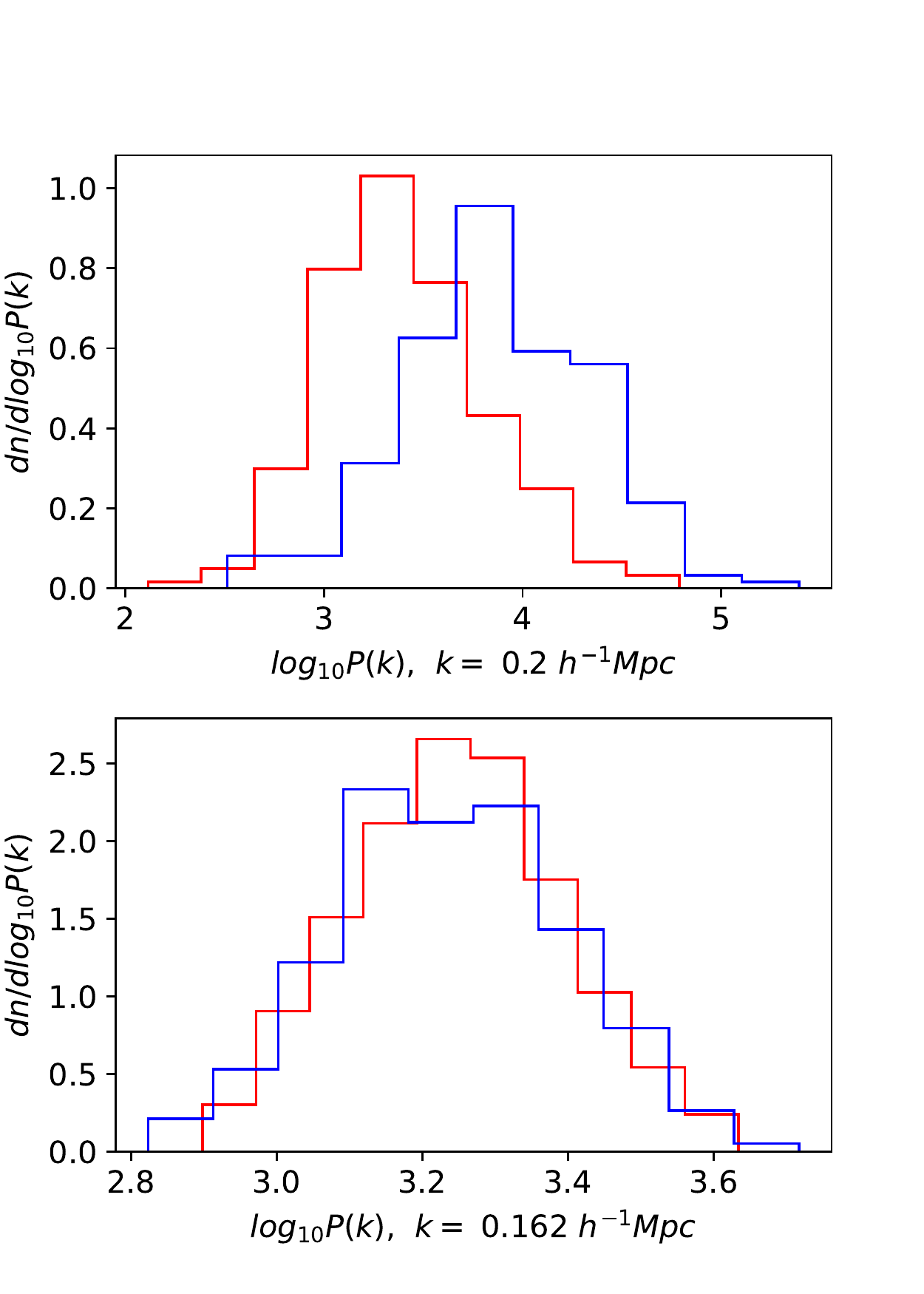}
     \end{adjustwidth}
    \protect\caption{Distributions of the power of a small (top) and large (bottom) scale, for the 250 spheres with the highest (red) and lowest (blue) Mach numbers $\Ml$. The power on large scales is measured  in a cube of side $120 \, \mathrm{Mpc\,h^{-1}}$, around each sphere. The power of small-scale fluctuations is clearly correlated with the local CMN $\Ml$, but the large scale fluctuations are indistinguishable, suggesting the driver for the difference in $\Ml$ on a given scale is the velocity dispersion of halos on that scale.}
     \label{Pk_dist}
 \end{figure}

 Another way of looking at the relation between $\Ml$ and the environment is to look at the \textit{local} power spectrum. We calculate the  power spectrum of the density field in each sphere (corresponding to scales smaller than the radius)  along with the power spectrum in a cube of $(120 \, \mathrm{h^{-1}Mpc)^3}$ centered around the sphere, to measure the fluctuations on scales larger than the sphere.  Figure \ref{Pk_dist} shows, for  250 spheres with the lowest local CMN $\Ml \sim 0.5$ (blue) and highest $\Ml \sim 3.5$ (red), the distribution of the value of the power spectrum for a typical large scale mode  $k=0.2 \, \mathrm{h/Mpc}$ (top panel)  and small scale mode $k=0.162 \, \mathrm{h/Mpc}$  (bottom panel). While the two histograms are almost identical for large sales, the distributions for  small-scale fluctuations are clearly distinct, the power is typically larger  for lower Mach numbers, as they typically have larger $\sigma$. The average large scale fluctuations do not seem to  affect  $\Ml$ as strongly as small scale modes, suggesting $\Ml$ is a good probe of the small scale power. These results show the same trend for all small and large modes we measured.

 \subsection{The rank ordered global Mach number}\label{rank}
As we have shown in the previous section environmental effects are not easy discernible by measuring $\Ml$ for an individual sphere as they show large scatters around general trends. To overcome this limitation we here propose to enhance any potential signal in the data by stacking spheres of similar $\Ml$ and using them to calculate a {\it rank ordered} global CMN via $\Mgr \equiv \left< u \right> /\left< \sigma \right>  $, where the average is taken over 50 spheres that are chosen to have similar local Mach numbers $\Ml$. The halo properties associated with these stacked spheres are calculated as the average of their value in each sphere of the stack. We focus on this ranked ordered Mach number instead of e.g. the average local Mach number, as $\Mgr$  can be  connected to cosmology and  the fluctuations of the power spectrum (see equations \ref{eq:u2} \&  \ref{eq:sigma2}).

We use  $4000$ non-overlapping spheres of the catalogs and sort them based on their local $\Ml$. Then, the $50$ spheres with the highest local Mach numbers are selected and used to calculate one global $\Mgr$, which is expected to be high. Another global $\Mgr$ is then calculated using the next $50$ spheres, and so on. Since spheres with a large $\Ml$ have a  relatively large bulk flow and a small velocity dispersion we expect $\Mgr$ calculated based on these $\Ml$ to be large as well. By grouping spheres in this manner, the $\Mgr$ distribution covers a larger range in $\delta$ than the $\Mg$ distribution, as we force the creation of a number of extremely low and high $\Mgr$ values that would only rarely appear for $\Mg$ if the selection of spheres was kept fully random in the averaging process.

The red points of Figure \ref{Mudispvsrho} show the rank ordered Mach number $\Mgr$, $u$, and $\sigma$, obtained by the selection mentioned above. 
The linear relation between $\sigma$ and the local overdensity remains, but a new tight relation appears between $u$ and $\delta$. As a result, $\Mgr$ is now strongly correlated with $\delta$  and can be reasonably well fitted by a simple relation of the form $\Mgr = \frac{1}{\alpha + \beta \delta}$, with $\alpha = 0.895$ and $\beta = 2.68$.\footnote{ Note that the fit parameter values will depend on selection criteria. As shown in Appendix \ref{app:sampling}, changing the halo mass range over which the CMN is calculated e.g. induces a small systematic effect which alter their  value by a few percents. However, we do not see this effect to change the general form of the  observed relation.  }

The red points in Figure \ref{Mvsprop} show the relations between $\Mgr$ and  halo properties. Again, the selection process and the grouping reveal a number of correlations that did not clearly appear with $\Ml$ (see Figure \ref{Mvsprop} blue points). The relation between $\Mgr$ and the total halo mass $M_{{\mathrm vir,tot}}$ in the spheres, which is directly linked to $\delta$ is becoming much clearer. Because of the relation between $M_{{\mathrm vir,tot}}$ and other halo properties, correlations with $\Mgr$ emerge. It is interesting to note that, when the global CMNs $\Mg$ are calculated using random spheres, they are likely to lie somewhere between $\sim 1$ and $\sim 1.5$, where the scatter is too large and  erases most of the trends in the Figure. One particular interesting feature is the fact that $\Mgr$ increases with the average scale factor at which the last major merger took place.  We observe later mergers in regions with a high Mach number. This would suggest that regions with a high Mach number (and therefore a low density) evolve more slowly than denser regions  as is expected in the $\Lambda$CDM paradigm.
We provide  fits for the relations between $\Mgr$ and the halo properties in table \ref{tab:fit_prop}.

The red points in Figure \ref{allvsrhos} show again results for $\Mgr$. The trends for $\Ml$ with $u$, and $\sigma$ are recovered although with much tighter scatter for $\Mgr$. However, a new clear anti-correlation between the environmental density fluctuation and the local density  appears. This explains the trend seen in the red points of the middle panel of Figure \ref{Mudispvsrho}.  Since $u$ increases with the environmental density fluctuation, which appears to decrease with $\delta$, $u$ now decreases with $\delta$. As a result, $\Mgr$ is now a tight linear function of these different features of the density field.   As such, it can be used as a proxy to measure them and extend the use of the CMN as a probe of structure formation in the Universe. 

We highlight the impact of the rank ordering in Figure \ref{rhoscolor} which shows the environmental density fluctuation against the local density for the 4000 spheres of radius $20 \Mpc$, presented in Figure \ref{allvsrhos},but color coded to represent the two groups of 50 spheres used to calculate the highest (red) and lowest (blue) global Mach numbers. The side panels show the histograms of $\delta$ and the EDF for each group. 
The trend shown in red on the first panel of Figure \ref{allvsrhos} is highlighted here by the selection process. Groups of high $\Mgr$  typically correspond to high EDF and low local densities, while lower $\Mgr$ lie lower on the figure, indicating a lower EDF.
The average of the histograms changes with $\Mgr$, and the distributions of the bottom panel get wider as $\Mgr$ decreases. However, while there is a relation between the CMN and enclosed mass, we note that the shape of the mass function does not change significantly as $\Ml$ or $\Mgr$ increases.

\begin{figure}
    \begin{adjustwidth}{-0.6cm}{0.0cm}
    \centering
    \includegraphics[scale=0.65]{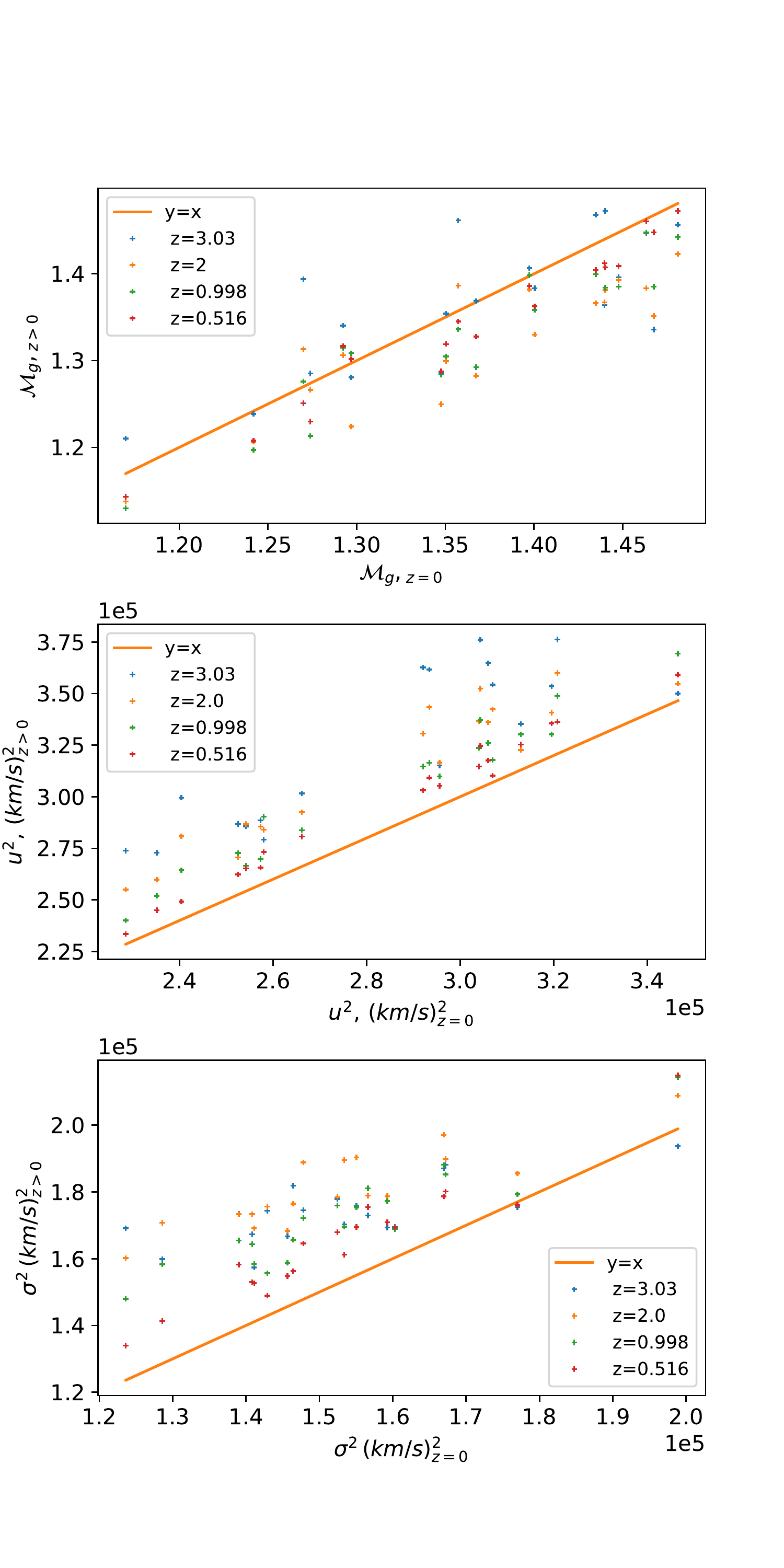}
    \end{adjustwidth}
   \protect\caption{ $\Mgr,\,u^2,\,\sigma^2$ for redshift 0 to 3, for rank ordered groups of spheres of $20 \Mpc$. The same spheres were followed as $z$ increases. The values for the bulk flow and the dispersion have been linearly extrapolated to z=0. The high-z bulk flow is in good agreement with estimates at z=0, but the dispersions significantly differ at higher redshift, suggesting that we deviate from the linear regime at z=0 even after the smoothing.  }   
    \label{gallvsz}
\end{figure}

To investigate further the impact of cosmic evolution on the CMN we use the rank ordered $\Mgr$ and follow the exact same co-moving spheres back in time and calculate their properties.
  Figure \ref{gallvsz} show how $\Mgr$, and the corresponding linearly extrapolated $\sigma$ and $u$ evolve as $z$ increases.    At $z=2$, the difference in $\Mgr$ can be as large as $20 \% $ for high-$\Mgr$ groups, but is close to $5 \%$ on average. In addition, the values of the linearly extrapolated bulk flow are consistent with the results at $z = 0$, but linearly extrapolating $\sigma^2$ results in an overestimate of $\sim 20 \%$ at $z = 3$. This suggests that, although we smooth over non-linearities this does not fully remove them and that they still can impact results slightly, which complements the results shown on Figure \ref{Mvsz}. As $z$ decreases, non linearities develop. This mostly impacts the small-scale velocity fluctuations, and the difference between $\sigma(z>0)$ and $\sigma(z=0)$ increases as $z$ increases.

\begin{table}
    \centering
    \begin{tabular}{c|c|c}
    \hline
       property  & A & B  \\
       \hline
        $v_{\mathrm{rms}}$ & -0.073 & 20.00 \\
        Spin & -357.0  & 15.10 \\
        ${\mathrm{log}}_{10}(M_{\mathrm{vir,tot}})$ & -1.774  & 27.66 \\
         $R_{\mathrm{vir}}$ &-0.047   & 16.67 \\
         scale factor of last major merger & 34.71 & -17.68\\
         $v_{\mathrm{max}}$ & -0.077 & 22.07 \\
        \hline

         \end{tabular}

   \protect\caption{Parameters used to fit $\Mgr$ to the halo properties with a linear function : $\Mgr = A \times \mathrm{property} + B$ }
    \label{tab:fit_prop}
\end{table}

\begin{comment}

\begin{figure}
\begin{adjustwidth}{-0.5cm}{0cm}
    \centering
    \includegraphics[scale=0.6]{figs_cap4/MF_spheres_av_20Mpc.pdf}
\end{adjustwidth}
   \protect\caption{\protect\footnotesize{Average  probability density function (top) and distribution (bottom) of the mass of the halos in the 10 groups of 50 spheres with the highest (thick red line) and lowest (thick blue line) $\Mgr$. There are more halos in groups with small   $\Mgr$ across all mass ranges, which is consistent with the relation between  $\Mgr$ and the total halo mass. However, the distributions show very similar shapes once normalized by mass. } }
    \label{MFglobal}
\end{figure}

\end{comment}

\section{Discussion and conclusion}
\label{sec::conclusion}
In this work, we revisit the role of the Cosmic Mach Numberas a probe of the environment of halos introducing new measures.
We have used a large scale cosmological N-body simulation from the Legacy project to calculate global Mach numbers for scales between $10$ and $100 \Mpc$,  comparing the theoretical predictions using the  input power spectrum of the simulation and the power spectrum measured from the matter distribution in the  simulation against Mach numbers calculated using halos found in the simulation. All these methods show  good agreement when the density field is smoothed over non-linear scales of $5 \Mpc$. We also highlight that smoothing the field and removing non-linearities is essential to achieve good agreement in the  calculation of the CMN.

Including all dark matter particles in a region, and not just dark matter halos, improves the estimates of the CMN compared to theoretical expectation  from $17 \%$ to $10 \%$  relative error and suggest intrinsic errors for estimates from galaxy surveys of the same level. 
We show that  simulations are intrinsically limited in the accuracy they can achieve for the CMN based on the range of modes they include. Any simulation used to calculate Mach numbers with accuracy up to $\sim 10 \% $ compared to theoretical predictions must include modes as large as $\sim 1000 \Mpc$ and as small as $\sim 5 \Mpc$ to accurately resolve all relevant scales.  Present state-of-the-art simulations such as Eagle (\cite{eagle1}, \cite{eagle2}) or Illustris (\cite{TNG}) do resolve small scale-fluctuations, but consist of simulation volumes of a few hundred $\Mpc$, too small to include relevant large sale modes. Simulated CMN computed using these simulation may be $\sim 5\%$ too small at scales of $\sim 10 \Mpc$, and up to $\sim 30 \%$ too small at larger scales ($\sim 100 \Mpc$), compared to expected results.

In addition, we study the relation between local Mach numbers and the properties of both the density field and the halos of a region. First, we corroborate the findings of previous studies, that  the velocity dispersion strongly depends on the overdensity of the region and the bulk flow does not, which results in  the local Mach number being a decreasing function of overdensity. Since  dark matter halos act as tracers of the density field, it weakly correlates with the total halo mass within a region as well. We also show that the bulk flow of a sphere is positively correlated with the density difference between two diametrically opposed adjacent larger spheres $\eta_{\rm{EDF}}$, which acts as a measure of the density gradient of the region. The local CMN $\Ml$ thus provide a proxy of   the relative strength of the density fluctuation measure $\eta_{\rm{EDF}}$ in contrast to the local density. The reason for these correlations seems to be two-fold. First, the relation between $u$ and  $\eta_{\rm{EDF}}$ is clear, halos are drawn towards dense regions, especially if this attraction is not compensated by the action of another dense region in the opposite direction, hence this is why $u$ is correlated with the density gradient. 
Concerning $\sigma$, the velocity dispersion is conceptually similar to the pair-wise velocity dispersion $\sigma_{12}$, which is in turn closely linked to the local density by the Cosmic Virial Theorem. The regions we study are not virialized, but they are overdensities on their way towards eventual collapse and virialization. This suggests that $\sigma$ and $\sigma_{12}$ would behave in similar ways. We also show that the local Mach numbers are qualitatively consistent with the power spectra in that region.

While the local Mach number shows a correlation that has a large scatter with $\eta_{\rm{EDF}}$ it does not show strong correlations with local environmental measures or average halo properties. Therefore, we introduce a new quantity, the rank ordered global Mach number $\Mgr$ by grouping spheres of similar local Mach numbers together and use them to calculate the global cosmic Mach number instead of using a random sample of regions. 

 $\Mgr$ shows strong empirical correlations with many halo properties, such as total mass, average radius and velocity dispersion, for which we provide fits in Section \ref{tab:fit_prop}, as well as with the local density and the environmental density fluctuation.
 Grouping the spheres in that manner also reveals a correlation between the enclosed density and $\eta_{\rm{EDF}}$. 

  The impact of non-linearities can be seen in the deviation of $\Mg$ from expectations based on the linear extrapolated power spectrum. While at low redshift we find deviations from the theoretical predicted CMNs for spheres with $r> 40 $ Mpc these vanish at $z\ge 3$ when the power spectrum on all scales follows the linearly extrapolated one.   
As shown in appendix \ref{app:sampling}, our findings also suggest that focusing solely on massive objects does not significantly alter the derived global Mach numbers. We studied the effect of focusing on halos more massive than $10^{12.5} \, M_\odot$ and $10^{13} \, M_\odot$, and find the average CMN to be  within 2 \% and 8 \%, respectively,  of the value obtained using all halos above $ 10^{12.2} M_{\odot} $.

The results presented in this paper are the first part of a study exploring the suitability of the CMN as an environment probe. While we here focus on the exploration of N-body simulations, further studies will focus on the observational viability as well as on the comparison to other common environment measures. However,
 our results suggest that from an observational point of view the most efficient strategy to derive environmental information using the CMN  on a given scale would be to group regions of similar local CMN $\Ml$  together for the calculation of their rank ordered $\Mgr$ and use such grouped spheres to investigate environmental effects of structure formation on the galaxy population. The grouping of spheres with similar $\Ml$ will allow to enhance underlying correlations and allow the efficient use of peculiar velocity surveys such as 6DFGS, the SDSS peculiar velocity catalogs. Studies based on the CMN  will serve as a complementary approach to classic environment measures based on e.g. number densities and have the potential to be more robust given the weak dependence on survey depths compared to number density studies. In a future study we plan to investigate correlations between the CMN and classic environmental measure in more detail, as well as extend this work to galaxies instead of halos.

\begin{section}{Acknowledgements}
The Legacy simulations presented here were run on ARCHER and Cirrus hosted by EPCC. SK is grateful for support from the UK STFC via grant ST/V000594/1.'
\end{section}
\begin{section}{Data availability}
The data underlying this article will be shared on reasonable request to the corresponding author.
\end{section}
\appendix{

 \section{Impact of smoothing scales and averaging procedure }
 \label{app:smooth}
 %\adg{RM : Calculating a single value of $\Mg$ with more spheres in a group allow us to sample the density field a bit better, but takes also longer. We did not observe a significant change in CMN by using more than a few dozen spheres in a each group, so we chose to use 50 in order to limit the computation time. As the radius increases, both the average and width of the distributions decrease.} \adr{SK: I would recommend to add a plot showing how the distribution changes as the number spheres increases in the Appendix} 

\begin{figure*}%[h!]
    \begin{adjustwidth}{-1.8cm}{-1.3cm}
    \centering
    \includegraphics[scale=0.8]{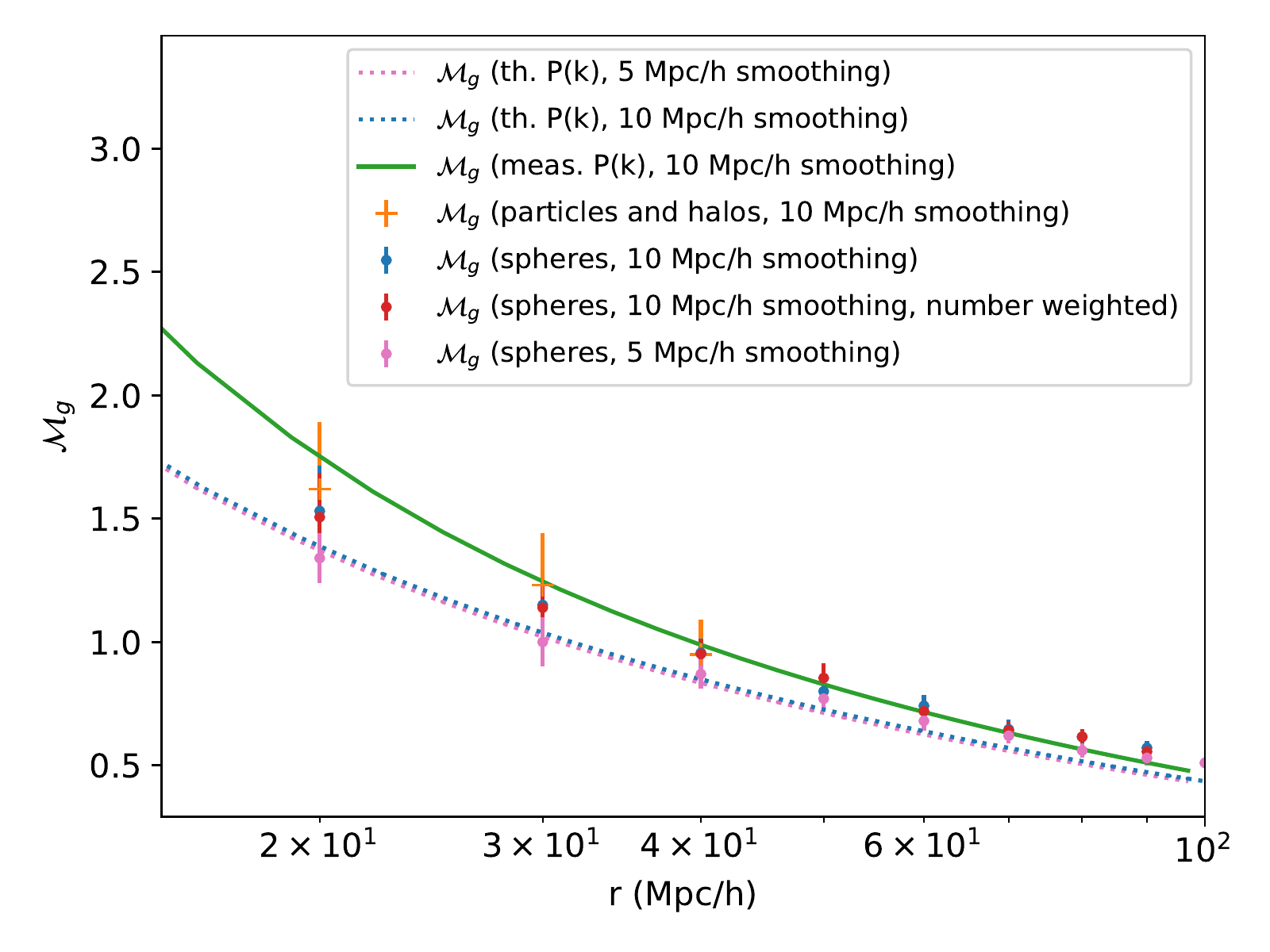}
    \end{adjustwidth}
    \caption{\small{$\Mg$ vs $R$, for different smoothing lengths and averaging methods. Using the masses as weights does not seem to significantly impact the results, as the blue and red points overlap. The Mach numbers calculated with halos are consistent with theoretical predictions when the smoothing length $r_s$ is set to $5 \Mpc$, but lie consistently lower when $r_s$ is increased to $10 \Mpc$}. }
    \label{Mvsr_rs10}
\end{figure*}

 We here discuss the impact of  smoothing non-linearities on various scales and the used averaging procedure on the predicted CMNs.
 Figure \ref{Mvsr_rs10} shows measures of the Mach number following the same smoothing procedure  with a smoothing length of  $10 \Mpc$ instead of the fiducial $5 \Mpc$. We also show results calculated with a smoothing length of $10 \Mpc$ and where the halos are averaged by number (we use $w_i = 1$ in Equation 10 ).  
 While the theoretical and measured Mach numbers are in good agreement at small scales when the field is smoothed over $5 \Mpc$, a significant discrepancy appears when using a $10 \Mpc$ smoothing length.  The averaging method does not seem to matter, as the results from the mass-weighted and number-weighted (see Equation (9)) procedures are almost identical. A possible explanation for the observed discrepancy is that halos are not perfect tracers of the density field. To test this, we perform the calculation of M using both the halos and all the dark matter particles that do not belong in a halo within a sphere. We here show the results calculated with 10 groups of 10 spheres. Including particles greatly increases the computation time, so the sample size was decreased.  This seems to improve our results at low scale (even though it would benefit from an increased sample size), as we more efficiently capture the density modes with the particles. The amplitude of the effect seems to increase with the smoothing scale, as the particles are not needed for consistent results with a $5 \Mpc$ smoothing scale.   Indeed, by using the particles, we can measure Mach numbers that are consistent with those predicted by the measured power spectrum. However, in this study, we calculate CMNs with dark matter halos since observations focus on galaxies, which reside in halos. We note that there is a $\sim 10 \%$ discrepancy between the results using halos only and those also including particles at scales $\lesssim 30 \Mpc$, for a smoothing length of $10 \Mpc$.

\section{Effects of sample selection}
\label{app:sampling}

\begin{figure}
    \begin{adjustwidth}{-1cm}{-0.8cm}
    \centering
    \includegraphics[scale=0.6]{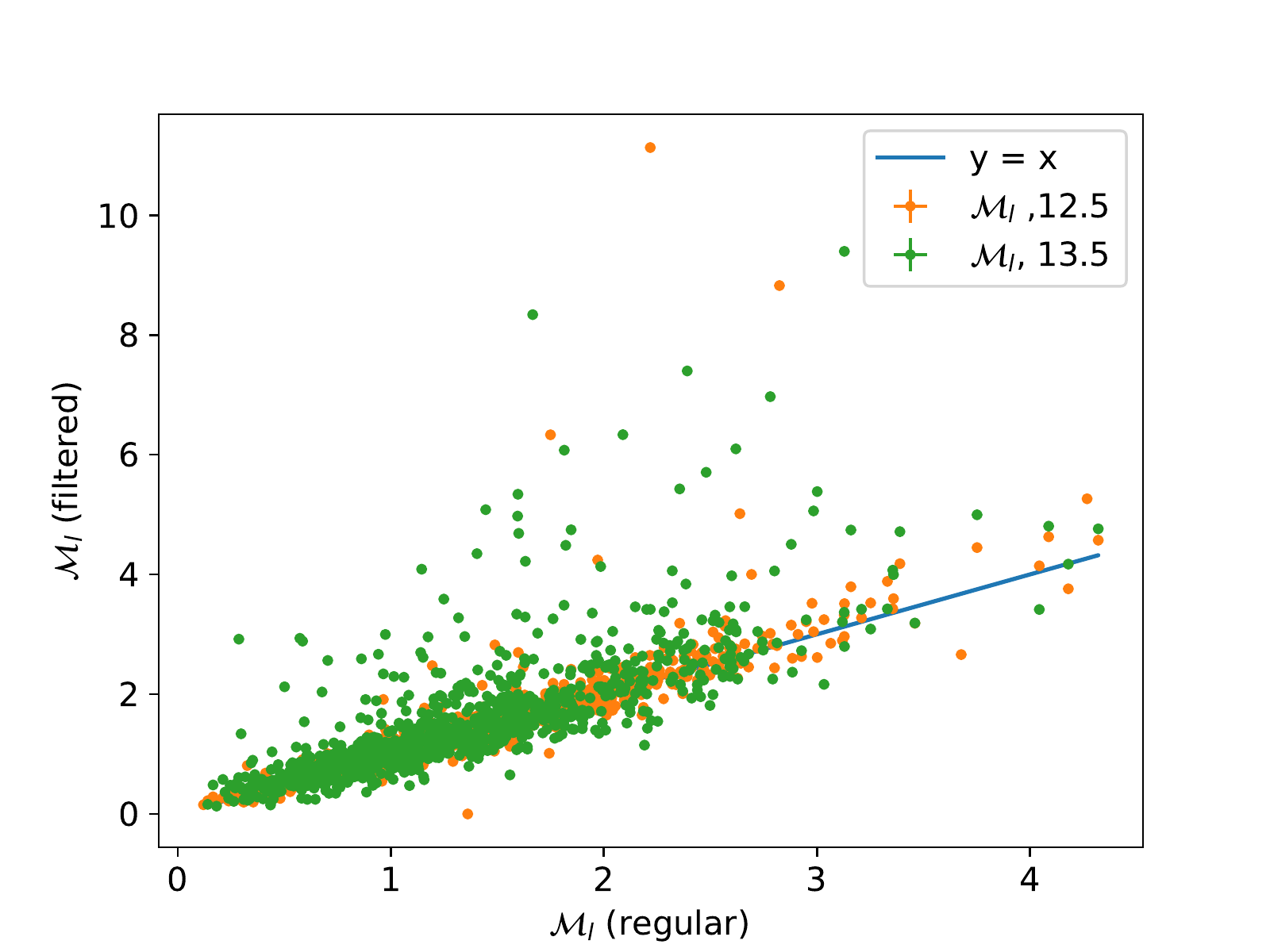}
    \end{adjustwidth}
    \caption{  \protect\footnotesize{The local Mach numbers $\Ml$ calculated using only halos more massive than $10^{12.5} \, M_\odot$ (orange points) or than $10^{13} \, M_\odot$ (green points) vs  $\Ml$ calculated with all  resolved halos. 
    The CMN calculated without the small halo are on average slightly larger than the actual CMN, with some rare rare points reaching significantly higher values. Although $\gtrsim 55 \%$ ($\gtrsim 85 \%$) of the halos are lighter than $10^{12.5} \, M_\odot$ ($10^{13} \, M_\odot$), the impact on the final global ranked Mach number is small  (2\% and 8\% on average) }.   }
    \label{MvMfilt}
\end{figure}

\begin{figure}
    \begin{adjustwidth}{-0.2cm}{-0.8cm}
    \centering
    \includegraphics[scale=0.6]{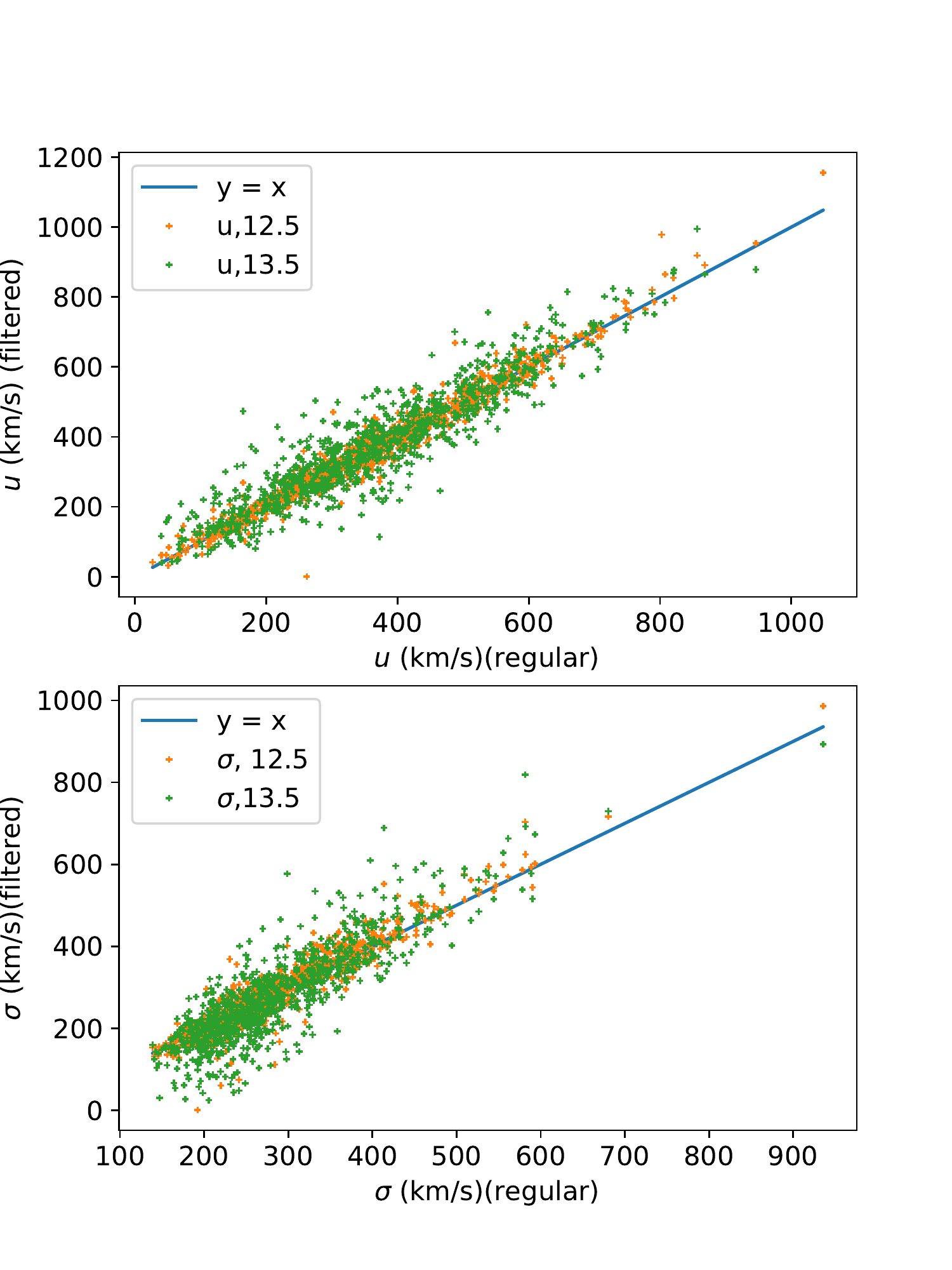}
    \end{adjustwidth}
    \caption{  \protect\footnotesize{Similar to \ref{MvMfilt}, but for $u$ (top) and $\sigma$ (bottom). On average, $u$ and $\sigma$ are not significantly affected by the mass filter, but the some dispersion are reduced to very small values, causing a large overestimate of the corresponding CMN.  }
    \label{usigvusigfilt}}
\end{figure}

Observational surveys have completeness limits based on their depth. To model the impact on the CMN we here apply mass cuts to halos in our simulation volume.  Figure \ref{MvMfilt} shows $\Ml$ calculated after filtering out all halos less massive than $10^{12.5} \, M_\odot$ (orange) or $10^{13} \, M_\odot$ (green) plotted against $\Ml$ without mass cut for the same group of spheres. We see that most Mach numbers remain virtually unchanged, which is  of interest for observational measures of $\Ml$. This also suggests that the small halos, (which on average account for $\sim 55 \%$ of the total number of halos in these $20 \Mpc$ spheres for the $10^{12.5} \, M_\odot$ filter and $\sim 85 \% $ for the $10^{13} \, M_\odot$ filter), are not the main drivers of $\Ml$. In the terms of the CMN massive halos are therefore accurate tracers of the underlying velocity fields and the associated matter power spectrum.  First, the average $\Ml$ is overestimated by 2\% (8\%) when using the $10^{12.5} \, M_\odot$ ($10^{13} \, M_\odot$) filter. Some CMNs reach extremely high values when calculated without the lighter halos : up to ten times the average. These come from spheres where the dispersion has been reduced to very small values (of the order of a dozen km/s), as only few, very massive objects remain in the region.

\begin{figure}
    \begin{adjustwidth}{-1cm}{-0.8cm}
    \centering
    \includegraphics[scale=0.6]{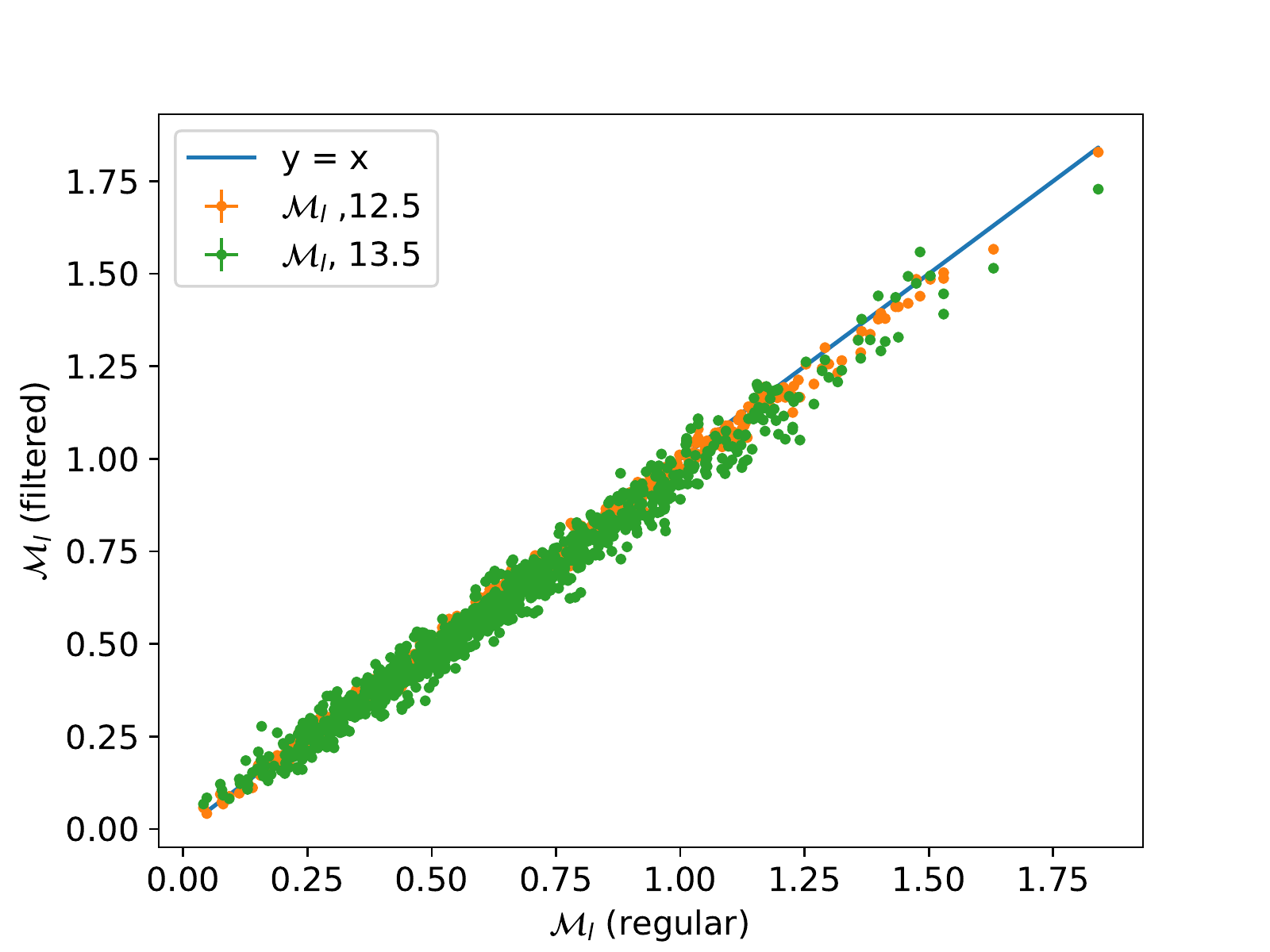}
    \end{adjustwidth}
    \caption{  \protect\footnotesize{Similar to \ref{MvMfilt}, but for spheres of $60 \Mpc$. With such large spheres, no extremely high $\Ml$ can be seen, but $\Ml$ tends to decrease by a few $\%$ as the lighter halos are removed.  }
    \label{MvMfilt_60}}
\end{figure}

\begin{figure}
    \begin{adjustwidth}{-1cm}{-0.8cm}
    \centering
    \includegraphics[scale=0.6]{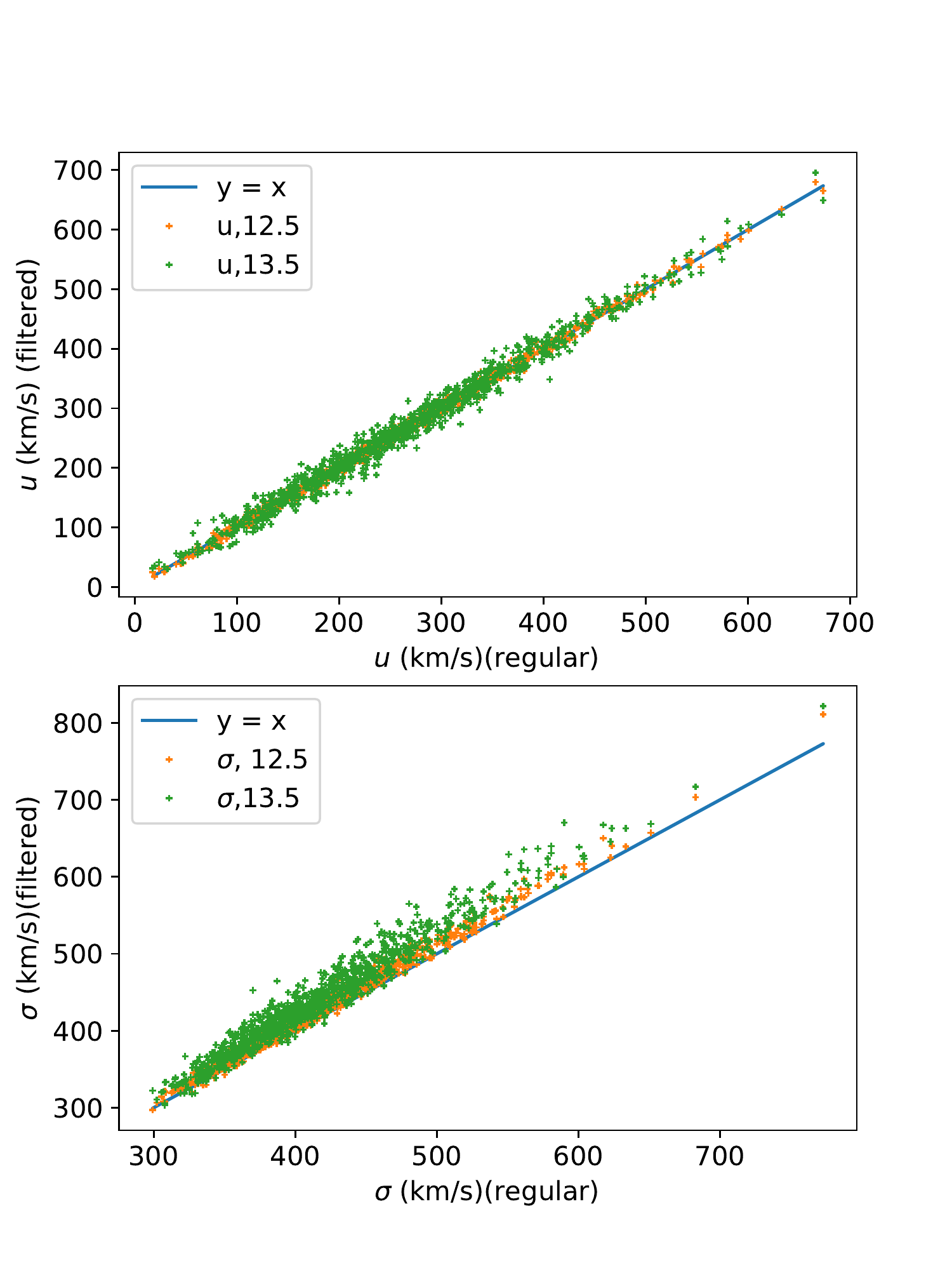}
    \end{adjustwidth}
    \caption{  \protect\footnotesize{Similar to \ref{usigvusigfilt}, but for spheres of $60 \Mpc$. $u$ suffers no systematic changes as the mass cut increases, but $\sigma$ tends to be overestimated by up to $10\%$ when only keeping halos more massive than $13.5  M_\odot$. } }
    \label{usigvusigfilt_60}
\end{figure}

 We note that this average systematic effect is smaller than the statistical dispersion around the average values of the CMN presented in Figure \ref{MvsR}, and therefore do not invalidate the conclusions drawn from it. 
Also, this systematic effect seem to disappear at larger radii, as we show in figures \ref{MvMfilt_60} and \ref{usigvusigfilt_60}. As few spheres are left with very small dispersion and few objects, no Mach numbers reach abnormally high values. However, they tend to decrease as the mass limit increases. Using only the most massive halos result in an overestimate of $\sim 10$ \% of the velocity dispersion, while no systematic change can be detected in $u$. This shows that smaller halos tend to more accurately follow the bulk flow formed by larger halos, which do not correctly represent the small scale fluctuations of the velocity field. Once more, this effect is roughly as strong as the observed scatter and is not likely to alter the conclusions of the previous sections.

}

%%%%%%%%%%%%%%%%%%%% REFERENCES %%%%%%%%%%%%%%%%%%

% The best way to enter references is to use BibTeX:

\bibliographystyle{mnras}
\bibliography{biblio} % if your bibtex file is called example.bib

\begin{thebibliography}{}
\makeatletter
\relax
\def\mn@urlcharsother{\let\do\@makeother \do\$\do\&\do\#\do\^\do\_\do\%\do\~}
\def\mn@doi{\begingroup\mn@urlcharsother \@ifnextchar [ {\mn@doi@}
  {\mn@doi@[]}}
\def\mn@doi@[#1]#2{\def\@tempa{#1}\ifx\@tempa\@empty \href
  {http://dx.doi.org/#2} {doi:#2}\else \href {http://dx.doi.org/#2} {#1}\fi
  \endgroup}
\def\mn@eprint#1#2{\mn@eprint@#1:#2::\@nil}
\def\mn@eprint@arXiv#1{\href {http://arxiv.org/abs/#1} {{\tt arXiv:#1}}}
\def\mn@eprint@dblp#1{\href {http://dblp.uni-trier.de/rec/bibtex/#1.xml}
  {dblp:#1}}
\def\mn@eprint@#1:#2:#3:#4\@nil{\def\@tempa {#1}\def\@tempb {#2}\def\@tempc
  {#3}\ifx \@tempc \@empty \let \@tempc \@tempb \let \@tempb \@tempa \fi \ifx
  \@tempb \@empty \def\@tempb {arXiv}\fi \@ifundefined
  {mn@eprint@\@tempb}{\@tempb:\@tempc}{\expandafter \expandafter \csname
  mn@eprint@\@tempb\endcsname \expandafter{\@tempc}}}

\bibitem[\protect\citeauthoryear{{Agarwal} \& {Feldman}}{{Agarwal} \&
  {Feldman}}{2013}]{Agarwal}
{Agarwal} S.,  {Feldman} H.~A.,  2013, \mn@doi [\mnras] {10.1093/mnras/stt464},
  \href {https://ui.adsabs.harvard.edu/abs/2013MNRAS.432..307A} {432, 307}

\bibitem[\protect\citeauthoryear{{Bartlett} \& {Blanchard}}{{Bartlett} \&
  {Blanchard}}{1996}]{CosmicVirialTheorem}
{Bartlett} J.~G.,  {Blanchard} A.,  1996, \aap, \href
  {https://ui.adsabs.harvard.edu/abs/1996A&A...307....1B} {307, 1}

\bibitem[\protect\citeauthoryear{{Behroozi}, {Wechsler}  \& {Wu}}{{Behroozi}
  et~al.}{2013}]{rockstar}
{Behroozi} P.~S.,  {Wechsler} R.~H.,   {Wu} H.-Y.,  2013, \mn@doi [\apj]
  {10.1088/0004-637X/762/2/109}, \href
  {https://ui.adsabs.harvard.edu/abs/2013ApJ...762..109B} {762, 109}

\bibitem[\protect\citeauthoryear{{Cen}, {Bahcall}  \& {Gramann}}{{Cen}
  et~al.}{1994}]{underestimatevel3}
{Cen} R.,  {Bahcall} N.~A.,   {Gramann} M.,  1994, \mn@doi [\apjl]
  {10.1086/187680}, \href
  {https://ui.adsabs.harvard.edu/abs/1994ApJ...437L..51C} {437, L51}

\bibitem[\protect\citeauthoryear{{Colberg}, {White}, {MacFarland}, {Jenkins},
  {Pearce}, {Frenk}, {Thomas}  \& {Couchman}}{{Colberg}
  et~al.}{2000}]{underestimatevel2}
{Colberg} J.~M.,  {White} S.~D.~M.,  {MacFarland} T.~J.,  {Jenkins} A.,
  {Pearce} F.~R.,  {Frenk} C.~S.,  {Thomas} P.~A.,   {Couchman} H.~M.~P.,
  2000, \mn@doi [\mnras] {10.1046/j.1365-8711.2000.03265.x}, \href
  {https://ui.adsabs.harvard.edu/abs/2000MNRAS.313..229C} {313, 229}

\bibitem[\protect\citeauthoryear{{Colombi}, {Jaffe}, {Novikov}  \&
  {Pichon}}{{Colombi} et~al.}{2009}]{errorbarspk2}
{Colombi} S.,  {Jaffe} A.,  {Novikov} D.,   {Pichon} C.,  2009, \mn@doi
  [\mnras] {10.1111/j.1365-2966.2008.14176.x}, \href
  {https://ui.adsabs.harvard.edu/abs/2009MNRAS.393..511C} {393, 511}

\bibitem[\protect\citeauthoryear{{Crain} et~al.,}{{Crain}
  et~al.}{2015}]{eagle1}
{Crain} R.~A.,  et~al., 2015, \mn@doi [\mnras] {10.1093/mnras/stv725}, \href
  {https://ui.adsabs.harvard.edu/abs/2015MNRAS.450.1937C} {450, 1937}

\bibitem[\protect\citeauthoryear{{Davis}, {Efstathiou}, {Frenk}  \&
  {White}}{{Davis} et~al.}{1985a}]{davishierarchy}
{Davis} M.,  {Efstathiou} G.,  {Frenk} C.~S.,   {White} S.~D.~M.,  1985a,
  \mn@doi [\apj] {10.1086/163168}, \href
  {https://ui.adsabs.harvard.edu/abs/1985ApJ...292..371D} {292, 371}

\bibitem[\protect\citeauthoryear{{Davis}, {Efstathiou}, {Frenk}  \&
  {White}}{{Davis} et~al.}{1985b}]{bardeenhierarchy}
{Davis} M.,  {Efstathiou} G.,  {Frenk} C.~S.,   {White} S.~D.~M.,  1985b,
  \mn@doi [\apj] {10.1086/163168}, \href
  {https://ui.adsabs.harvard.edu/abs/1985ApJ...292..371D} {292, 371}

\bibitem[\protect\citeauthoryear{{Feldman}, {Kaiser}  \& {Peacock}}{{Feldman}
  et~al.}{1994}]{errorbarspk1}
{Feldman} H.~A.,  {Kaiser} N.,   {Peacock} J.~A.,  1994, \mn@doi [\apj]
  {10.1086/174036}, \href
  {https://ui.adsabs.harvard.edu/abs/1994ApJ...426...23F} {426, 23}

\bibitem[\protect\citeauthoryear{{Hahn} \& {Abel}}{{Hahn} \&
  {Abel}}{2011}]{music}
{Hahn} O.,  {Abel} T.,  2011, \mn@doi [\mnras]
  {10.1111/j.1365-2966.2011.18820.x}, \href
  {https://ui.adsabs.harvard.edu/abs/2011MNRAS.415.2101H} {415, 2101}

\bibitem[\protect\citeauthoryear{Hand, Feng, Beutler, Li, Modi, Seljak  \&
  Slepian}{Hand et~al.}{2018}]{nbodykit}
Hand N.,  Feng Y.,  Beutler F.,  Li Y.,  Modi C.,  Seljak U.,   Slepian Z.,
  2018, \mn@doi [Astron. J.] {10.3847/1538-3881/aadae0}, 156, 160

\bibitem[\protect\citeauthoryear{{Hinshaw} et~al.,}{{Hinshaw}
  et~al.}{2013}]{WMAP9}
{Hinshaw} G.,  et~al., 2013, \mn@doi [\apjs] {10.1088/0067-0049/208/2/19},
  \href {https://ui.adsabs.harvard.edu/abs/2013ApJS..208...19H} {208, 19}

\bibitem[\protect\citeauthoryear{Kumar, Wang, Feldman  \& Watkins}{Kumar
  et~al.}{2015}]{Kumar2015GravitationalPW}
Kumar A.~N.,  Wang Y.,  Feldman H.~A.,   Watkins R.~J.,  2015, Bulletin of the
  American Physical Society, 2016

\bibitem[\protect\citeauthoryear{{Linder}}{{Linder}}{2005}]{Linder2005}
{Linder} E.~V.,  2005, \mn@doi [\prd] {10.1103/PhysRevD.72.043529}, \href
  {https://ui.adsabs.harvard.edu/abs/2005PhRvD..72d3529L} {72, 043529}

\bibitem[\protect\citeauthoryear{Ma, Ostriker  \& Zhao}{Ma
  et~al.}{2011}]{CosmicProbe}
Ma Y.-Z.,  Ostriker J.,   Zhao G.-B.,  2011, \mn@doi [Journal of Cosmology and
  Astroparticle Physics] {10.1088/1475-7516/2012/06/026}, 2012

\bibitem[\protect\citeauthoryear{{Nagamine}, {Ostriker}  \& {Cen}}{{Nagamine}
  et~al.}{2001}]{Moverdensity}
{Nagamine} K.,  {Ostriker} J.~P.,   {Cen} R.,  2001, \mn@doi [\apj]
  {10.1086/320966}, \href
  {https://ui.adsabs.harvard.edu/abs/2001ApJ...553..513N} {553, 513}

\bibitem[\protect\citeauthoryear{{Ostriker} \& {Suto}}{{Ostriker} \&
  {Suto}}{1990}]{1990ApJ...348..378O}
{Ostriker} J.~P.,  {Suto} Y.,  1990, \mn@doi [\apj] {10.1086/168247}, \href
  {https://ui.adsabs.harvard.edu/abs/1990ApJ...348..378O} {348, 378}

\bibitem[\protect\citeauthoryear{{Peebles}}{{Peebles}}{1980}]{peebles}
{Peebles} P.~J.~E.,  1980, {The large-scale structure of the universe}

\bibitem[\protect\citeauthoryear{{Planck Collaboration} et~al.,}{{Planck
  Collaboration} et~al.}{2020}]{Planck18}
{Planck Collaboration} et~al., 2020, \mn@doi [\aap]
  {10.1051/0004-6361/201833910}, \href
  {https://ui.adsabs.harvard.edu/abs/2020A&A...641A...6P} {641, A6}

\bibitem[\protect\citeauthoryear{{Schaye} et~al.,}{{Schaye}
  et~al.}{2015}]{eagle2}
{Schaye} J.,  et~al., 2015, \mn@doi [\mnras] {10.1093/mnras/stu2058}, \href
  {https://ui.adsabs.harvard.edu/abs/2015MNRAS.446..521S} {446, 521}

\bibitem[\protect\citeauthoryear{{Sheth} \& {Diaferio}}{{Sheth} \&
  {Diaferio}}{2001}]{underestimatevel1}
{Sheth} R.~K.,  {Diaferio} A.,  2001, Monthly Notices of the Royal Astronomical
  Society, 322, 901

\bibitem[\protect\citeauthoryear{Springel}{Springel}{2005}]{gadget}
Springel V.,  2005, \mn@doi [Mon. Not. Roy. Astron. Soc.]
  {10.1111/j.1365-2966.2005.09655.x}, 364, 1105

\bibitem[\protect\citeauthoryear{{Springel} et~al.,}{{Springel}
  et~al.}{2005}]{Springel}
{Springel} V.,  et~al., 2005, \mn@doi [\nat] {10.1038/nature03597}, \href
  {https://ui.adsabs.harvard.edu/abs/2005Natur.435..629S} {435, 629}

\bibitem[\protect\citeauthoryear{{Springel} et~al.,}{{Springel}
  et~al.}{2018}]{TNG}
{Springel} V.,  et~al., 2018, \mn@doi [\mnras] {10.1093/mnras/stx3304}, \href
  {https://ui.adsabs.harvard.edu/abs/2018MNRAS.475..676S} {475, 676}

\bibitem[\protect\citeauthoryear{{Springel}, {Pakmor}, {Zier}  \&
  {Reinecke}}{{Springel} et~al.}{2020}]{gadget4}
{Springel} V.,  {Pakmor} R.,  {Zier} O.,   {Reinecke} M.,  2020, arXiv
  e-prints, \href {https://ui.adsabs.harvard.edu/abs/2020arXiv201003567S} {p.
  arXiv:2010.03567}

\bibitem[\protect\citeauthoryear{{Strauss}, {Cen}  \& {Ostriker}}{{Strauss}
  et~al.}{1993}]{1993ApJ...408..389S}
{Strauss} M.~A.,  {Cen} R.,   {Ostriker} J.~P.,  1993, \mn@doi [\apj]
  {10.1086/172596}, \href
  {https://ui.adsabs.harvard.edu/abs/1993ApJ...408..389S} {408, 389}

\bibitem[\protect\citeauthoryear{Strauss, Ostriker  \& Cen}{Strauss
  et~al.}{1998}]{pairwise_sigma}
Strauss M.~A.,  Ostriker J.~P.,   Cen R.,  1998, \mn@doi [Astrophys. J.]
  {10.1086/305211}, 494, 20

\bibitem[\protect\citeauthoryear{{Suto}, {Cen}  \& {Ostriker}}{{Suto}
  et~al.}{1992}]{1992ApJ...395....1S}
{Suto} Y.,  {Cen} R.,   {Ostriker} J.~P.,  1992, \mn@doi [\apj]
  {10.1086/171626}, \href
  {https://ui.adsabs.harvard.edu/abs/1992ApJ...395....1S} {395, 1}

\bibitem[\protect\citeauthoryear{{Tinker}, {Kravtsov}, {Klypin}, {Abazajian},
  {Warren}, {Yepes}, {Gottl{\"o}ber}  \& {Holz}}{{Tinker}
  et~al.}{2008}]{TinkerMF}
{Tinker} J.,  {Kravtsov} A.~V.,  {Klypin} A.,  {Abazajian} K.,  {Warren} M.,
  {Yepes} G.,  {Gottl{\"o}ber} S.,   {Holz} D.~E.,  2008, \mn@doi [\apj]
  {10.1086/591439}, \href
  {https://ui.adsabs.harvard.edu/abs/2008ApJ...688..709T} {688, 709}

\bibitem[\protect\citeauthoryear{{Turner}}{{Turner}}{1997}]{caseLCDM}
{Turner} M.~S.,  1997, arXiv e-prints, \href
  {https://ui.adsabs.harvard.edu/abs/1997astro.ph..3161T} {pp
  astro--ph/9703161}

\bibitem[\protect\citeauthoryear{{Vikhlinin} et~al.,}{{Vikhlinin}
  et~al.}{2009}]{clustercosmo}
{Vikhlinin} A.,  et~al., 2009, \mn@doi [\apj] {10.1088/0004-637X/692/2/1060},
  \href {https://ui.adsabs.harvard.edu/abs/2009ApJ...692.1060V} {692, 1060}

\bibitem[\protect\citeauthoryear{Weinberg}{Weinberg}{2008}]{Cosmology}
Weinberg S.,  2008, Cosmology.
Oxford University Press

\makeatother
\end{thebibliography}

% Alternatively you could enter them by hand, like this:
% This method is tedious and prone to error if you have lots of references
%\begin{thebibliography}{99}
%\bibitem[\protect\citeauthoryear{Author}{2012}]{Author2012}
%Author A.~N., 2013, Journal of Improbable Astronomy, 1, 1
%\bibitem[\protect\citeauthoryear{Others}{2013}]{Others2013}
%Others S., 2012, Journal of Interesting Stuff, 17, 198
%\end{thebibliography}

%%%%%%%%%%%%%%%%%%%%%%%%%%%%%%%%%%%%%%%%%%%%%%%%%%

%%%%%%%%%%%%%%%%% APPENDICES %%%%%%%%%%%%%%%%%%%%%

%%%%%%%%%%%%%%%%%%%%%%%%%%%%%%%%%%%%%%%%%%%%%%%%%%

% Don't change these lines
\bsp	% typesetting comment
\label{lastpage}
\end{document}